\documentclass[12pt,preprint]{aastex}
\def\chandra{{\it Chandra}}
\def\xmm{{\it XMM-Newton}}
\def\suzaku{{\it Suzaku}}

\def\deg{$^{\circ}$}

\newcommand{\Msun}{\hbox{$\rm\thinspace M_{\odot}$}}
\newcommand{\ls}
{\mathrel{\hbox{\rlap{\hbox{\lower4pt\hbox{$\sim$}}}\hbox{$<$}}}}
\newcommand{\gs}
{\mathrel{\hbox{\rlap{\hbox{\lower4pt\hbox{$\sim$}}}\hbox{$>$}}}}
\def\Msun{\hbox{$\rm ~M_{\odot}$}}
\def\3c{3C~445}

\begin{document}

\title{Chandra High Resolution Spectroscopy of the Circumnuclear 
Matter in the Broad Line Radio Galaxy, 3C\,445}

\author{J. N. Reeves}
\affil{Astrophysics Group, School of Physical \& Geographical Sciences, Keele
University, Keele, Staffordshire ST5 5BG, UK; e-mail jnr@astro.keele.ac.uk}
\affil{CSST, University of Maryland Baltimore County, 1000 Hilltop Circle, 
Baltimore, MD 21250, USA}

\author{J. Gofford}
\affil{Astrophysics Group, School of Physical \& Geographical Sciences, Keele
University, Keele, Staffordshire ST5 5BG, UK}

\author{V. Braito}
\affil{University of Leicester, Department of Physics \& Astronomy,
University Road, Leicester LE1 7RH, UK}

\author{R. Sambruna}
\affil{Code 662, NASA Goddard Space Flight Center, Greenbelt, MD 20771, USA}

%\author{J. Gofford et al.}
%\affil{Astrophysics Group, School of Physical \& Geographical Sciences, Keele
%University, Keele, Staffordshire ST5 5BG, UK}

%\author{Michael Eracleous}
%\affil{Department of Astronomy \& Astrophysics and Center for Gravitational 
%Wave Physics, The Pennsylvania State University, 
%525 Davey Laboratory, University Park, PA 16802, USA}

\clearpage

\begin{abstract}

We present evidence for X-ray line emitting and absorbing gas in the nucleus
of the Broad-Line Radio Galaxy (BLRG), 3C\,445. A 200\,ks
Chandra LETG observation of 3C\,445 reveals the
presence of several highly ionized emission lines in the soft X-ray
spectrum, primarily from the He and H-like ions of O, Ne, Mg and Si. 
Radiative recombination emission is detected from O\,\textsc{vii} and 
O\,\textsc{viii}, indicating that the emitting gas is photoionized. 
The He-like emission appears to be resolved 
into forbidden and intercombination line components, 
which implies a high density of $>10^{10}$\,cm$^{-3}$, while the 
Oxygen lines 
are velocity broadened with a mean width of $\sim 2600$\,km\,s$^{-1}$ (FWHM). 
The density and widths of the ionized lines indicate an origin 
of the gas on sub-parsec scales in the Broad Line Region (BLR).
The X-ray continuum of 3C\,445 is heavily obscured 
either by a partial coverer or by a photoionized 
absorber of
column density $N_{\rm H}=2\times10^{23}$\,cm$^{-2}$ and ionization 
parameter $\log \xi=1.4$\,erg\,cm\,s$^{-1}$.
However the view of the 
X-ray line emission is unobscured, which requires the absorber to be located 
at radii well within any parsec scale molecular torus. Instead we suggest 
that the 
X-ray absorber in 3C\,445 may be associated with an outflowing, but clumpy 
accretion disk wind, 
with an observed outflow velocity of $\sim 10000$\,km\,s$^{-1}$. 
%The mechanical 
%output of the wind is estimated to be close to $10^{44}$\,erg\,s$^{-1}$, 
%similar to the X-ray luminosity of 3C\,445.

\end{abstract}


\section{Introduction}

The X-ray emission from AGN is a powerful tool to investigate the
structure and physical conditions of the matter in the proximity of
the central supermassive black hole. Sensitive X-ray spectroscopy has
been very successful in disentangling the contributions from warm and
cold matter in AGN; e.g. see \citet{turner09} for a review. 
At soft X-ray energies more than 50\% of nearby 
Seyfert 1s exhibit complex intrinsic absorption and/or emission lines
suggesting the presence of photoionized gas (\citealt{crenshaw03}; \citealt{blustin05}; \citealt{mckernan07}), which may contain 
a significant fraction of the accreting
mass. This is observed in the form of X-ray absorption in type 1 AGN, 
otherwise known as "warm absorbers", which have typical outflow velocities 
in the UV and X-ray band of 100--1000\,km\,s$^{-1}$ \citep{crenshaw03}.
The same absorbing gas is thought to
be responsible for the soft X-ray emission lines observed in type-2
sources \citep{gb07, kink02, turner97}, 
which may be associated with parsec scale 
gas, photoionized by the inner central engine.
Furthermore in a handful of radio-quiet AGN, blueshifted absorption features have been observed 
with higher velocity shifts, through detections of resonance absorption lines in the iron 
K band, indicating an outflow from the nucleus with
quasi--relativistic velocities, $v/c \sim 0.1 $ (e.g. see \citealt{tombesi10} and references 
therein). 

%These so-called ultra--fast outflows would imply large mass 
%outflow rates (e.g. \citealt{reeves09}) and may be energetically significant in 
%terms of AGN feedback, driving out significant amounts of matter from the galaxy 
%and limiting the ultimate growth of the black hole and host bulge (e.g. \citealt{king03}; 
%\citealt{king10}).
%While the statistical significance of some of these fast outflows has been called into 
%question \citep{vaughan08}, a substantial number of 
%fast outflows have been confirmed 
%in a uniform sample of Seyferts measured by XMM-Newton \citep{tombesi10}.

%While the circumnuclear X-ray gas in Seyfert galaxies appear 
%to be well studied, the situation is less clear among the radio-loud AGN.
Until very recently, the 
general consensus from the X-ray spectra of radio-loud 
AGN was that, unlike their radio-quiet cousins, 
they contained little or no ionized gas in their nuclei.
%this could constitute a major difference with their radio-quiet cousins, where
%a forest of X-ray lines in emission and absorption have been detected in
%both type~1 and 2 sources (Turner \& Miller 2009).
Thus there appeared to be no evidence for ionized emitting or absorbing 
gas in the soft X-ray spectra of Broad Lined Radio Galaxies (BLRGs).
For instance a 120~ks \suzaku\ observation of 3C~120
showed a featureless continuum at soft X-rays, attributed to the radio
jet \citep{kataoka07}, while a featureless soft X-ray
continuum was observed in 3C~390.3 \citep{sambruna09}. 

Recent sensitive observations with \chandra, \xmm, and \suzaku\ are
subverting this view. Lines in emission and absorption have been detected
at soft X-rays in type~1 (Broad-Line, BLRGs) and in type~2
(Narrow-Line, NLRGs) Radio Galaxies, indicating large gas column
densities, of $N_{\rm H} = 10^{21} - 10^{23}$\,cm$^{-2}$ and a range of
ionization parameters, $\log\xi \sim 1-5\;$ergs~cm~s$^{-1}$
(\citealt {sambruna07}, \citealt{grandi07}, \citealt{piconcelli08}, \citealt{torresi09}, 
\citealt{torresi09}, \citealt{reeves09}, \citealt{torresi10}, \citealt{tombesi10b}).
Ionized soft X-ray emission lines have so far been detected in the BLRG 3C~445 
(\citealt{sambruna07}; \citealt{grandi07}) 
and in the NLRGs 3C\,234 \citep{piconcelli08} and 3C~33 \citep{torresi09}. 
Photoionized absorption lines, consistent with gas outflowing on parsec 
scales with velocities of hundreds of km\,s$^{-1}$, were detected for the 
first time with grating resolution X-ray spectra in the BLRG 3C\,382, 
with Chandra/HETG \citep{reeves09} and independently with XMM-Newton/RGS 
\citep{torresi10}. Interestingly, Suzaku observations
of BLRGs has also uncovered evidence at higher energies, at $7-9$\,keV in the iron K band, 
for fast outflowing gas with velocities $v_{\rm out}\sim 0.04-0.15c$,
carrying substantial masses and kinetic powers similar to the radio
jets \citep{tombesi10b}.  Thus there appears to be substantial 
ionized gas in the nuclei of radio-loud AGN, and this gas may be an 
energetically important component that needs to be accounted for in
models for accretion and jet formation.

Indeed, there are reasons to expect the presence of such a medium in
BLRGs and other radio-loud AGN. For example centrifugally-driven
winds, lifting matter off the disk's surface and channelling it down
the magnetic field, are a proposed scenario for the origin of
relativistic jets \citep{blandford82}; at
favorable orientations, these winds lead to observable discrete
absorption/emission features at soft X-rays \citep{konig94}.  
Jet formation models predict that the
relativistically moving plasma should be enveloped in a
sub-relativistic wind \citep{mckinney06}, with
velocities $\ls 0.1c$. Unification models
for radio-loud sources also postulate the presence of a warm,
scattering gas to explain type-2 sources (\citealt{antonucci93}; \citealt{urry95}).

\subsection{The Broad Lined Radio Galaxy 3C 445}

3C\,445 is a bright, nearby ($z=0.0562$, \citealt{hewitt91}; \citealt{erac04}) and luminous ($L_{\rm bol}\sim10^{45}$\,erg\,s$^{-1}$, \citealt{marchesini04})
radio galaxy with an FRII morphology \citep{kronberg86}. 3C\,445
appears lobe rather than core dominated \citep{morganti93} and is likely to be 
highly inclined with respect to the radio-jet axis, with an inclination angle of 
$\sim60-70$\deg\ (\citealt{erac98}; \citealt{sambruna07}). 
Based on its optical spectra it is classed as a BLRG, due to the presence 
of strong broad permitted lines in unpolarized light (\citealt{osterbrock76}; \citealt{crenshaw88}, 
\citealt{erac94}; \citealt{corbett98}). 
From its rather large Balmer decrement, 
the line of sight reddening towards 3C\,445 is $E_{B-V} \sim 1$, which for a Galactic dust to gas ratio 
suggests an absorbing column density of $N_{\rm H}\sim5\times10^{21}$\,cm$^{-2}$, an order 
of magnitude higher than the Galactic line of sight column \citep{dickey90}. 

3C\,445 is also a bright source in the X-ray band, having previously been detected by 
EXOSAT \citep{turner89}, Ginga \citep{pounds90}, ASCA \citep{sambruna98} and is 
also detected in the hard X-ray band (above 10 keV), with Beppo-SAX \citep{grandi06}, 
Swift/BAT \citep{tueller10} and most recently with Suzaku \citep{braito10}. Past observations 
with EXOSAT, Ginga and ASCA all indicated an absorbed X-ray spectrum, with a column density of 
$N_{\rm H}\sim 10^{23}$\,cm$^{-2}$, far in excess of the column density expected from the 
amount of reddening in the optical spectra of 3C\,445. A more recent short (15\,ks) XMM-Newton 
observation of 3C\,445 confirmed the absorbed nature of its X-ray emission, with the 
absorber either partially covering the AGN, or consisting of partially ionized material 
(\citealt{sambruna07}; \citealt{grandi07}). 
%with further tentative evidence in the iron K band for highly ionized 
%absorption from a possible outflow.

Most interestingly, the XMM-Newton observations suggested the presence of multiple highly ionized soft 
X-ray emission lines (\citealt{sambruna07}; \citealt{grandi07}), primarily from O, Mg and Si, with 
a spectrum somewhat reminiscent of those of Seyfert 2 galaxies \citep{gb07}. 
Given the relatively low exposure and lower resolution of the XMM-Newton EPIC-pn data below 2 keV, 
it was impossible to deduce the physical properties or location of the emitting gas, which 
was constrained to lie within $<5$\,kpc of the nucleus in 3C\,445 \citep{sambruna07}.  
A tentative detection of the 
O\,\textsc{vii} and O\,\textsc{viii} emission lines was made in the XMM-Newton RGS observations 
\citep{grandi07}, however the short exposure precluded a more detailed analysis of the 
soft X-ray line emitting gas.

In this paper we present direct evidence for the photoionized circumnuclear 
gas around the nucleus of 3C\,445, from high resolution spectroscopy with the Chandra LETG 
(Low Energy Transmission Grating, \citealt{brinkman2000}). In this much deeper 200\,ks exposure, we resolve 
multiple emission lines 
in the soft X-ray band from the high resolution LETG data. The higher quality 
of the data allows us to determine the properties and location of the emitting matter, 
which as we will subsequently show, is most likely to be emission from highly ionized gas 
associated with the BLR clouds in 3C\,445. The Chandra LETG spectrum also allows an 
accurate measurement of the properties of the absorbing gas towards 3C\,445, which appears 
to be outflowing with respect to the rest frame of 3C\,445.
A subsequent paper (Braito et al. 2010, in preparation) 
will discuss in detail the broad--band X-ray spectrum of 3C\,445  
observed with Suzaku and Swift. 

The organization of this paper is as follows. In \S~2 we describe the
\chandra\ data reduction and analysis; in \S~3 the
results of the spectral analysis \S4 the photoionization modeling of the 
spectrum; Discussion and Conclusions follow in
\S~5 and \S~6. Throughout this paper, a concordance cosmology with H$_0=71$ km\,s$^{-1}$ Mpc$^{-1}$, $\Omega_{\Lambda}$=0.73, and $\Omega_m$=0.27
\citep{spergel03} is adopted. Errors are quoted to 90\% confidence 
for 1 parameter of interest (i.e. $\Delta \chi^{2}$ or $\Delta C=2.71$).
All the spectral parameters in subsequent sections are quoted in the 
rest--frame of 3C\,445 ($z=0.0562$) unless otherwise stated.

\section{The Chandra LETG Observations} 

Chandra observed 3C\,445 with the LETG (Low Energy Transmission Grating) 
for a net exposure of 198\,ks
between 25 September -- 3 October 2009, with the ACIS-S detector in 
the focal plane. The zeroth order image of 3C\,445 at the aim-point of 
ACIS-S appears unresolved.
No significant source variability
was found during the observations, so the time--averaged data was used.
The $\pm1$ order spectra were summed
for the LETG respectively, along with their response files.  The 
resultant summed (background subtracted) first order
count rate for the LETG is $0.0318\pm0.0004$\,counts\,s$^{-1}$ over the energy 
range 0.5--9.0\,keV. 

\section{The Soft X-ray Spectrum of 3C\,445}

\subsection{Initial Continuum Modeling}

Initially, in order to parameterize the LETG spectrum of 
3C\,445 with simple continuum models, we binned the 
spectra to a minimum of 10 source counts per bin and apply 
the $\chi^{2}$ statistic. 
The continuum of the LETG spectrum from 3C\,445 can be very 
crudely parameterized by a broken--powerlaw model, as is shown in 
Figure 1. The spectrum at higher energies, above a break energy of 
$\sim 1.4$\,keV ($\sim9$\,\AA) is very hard, rising with energy (decreasing wavelength) with a 
photon index of $\Gamma=-0.64\pm0.03$. Below the break, the X-ray 
spectrum is softer, with $\Gamma=1.73\pm0.21$. This initial parameterization 
of the data provided a poor fit (with $\chi^{2}/{\rm dof}=568.5/181$, null 
hypothesis probability $P=1.3\times 10^{-41}$, where dof is the number of degrees of 
freedom in the fit). At soft X-ray energies, 
there is statistically significant scatter around the continuum due to the 
presence of likely emission lines, 
while at higher energies, residuals 
are present in the data, most notably emission and 
absorption in the iron K band between 
6--8\,keV ($\sim 2$\,\AA). The observed flux of 3C\,445 from 0.5--9\,keV is 
$8.84\times10^{-12}$\,ergs\,cm$^{-2}$\,s$^{-1}$.

In order to provide a more physical 
representation of the continuum, we fit the 
spectrum with a photoelectric absorption model of the form 
$F(E) = {\rm wabs} \times ({\rm zwabs}\times{\rm pow1} + {\rm pow2})$, 
where ${\rm pow1}$ represents 
an absorbed power--law, ${\rm pow2}$ the unabsorbed power--law continuum (i.e. 
absorbed only by the Galactic column density),
${\rm wabs}$ is the local Galactic line of sight 
absorber (where $N_{H, Gal}=4.6\times10^{20}$\,cm$^{-2}$ for 3C\,445, 
\citealt{dickey90}) and ${\rm zwabs}$ is the intrinsic absorber of column density $N_{H}$ towards
3C\,445, fitted in the rest--frame of the radio galaxy ($z=0.0562$). 
Solar abundances 
are assumed \citep{AG89}, while the neutral absorber uses cross-sections
of \citet{morrison83}. 

The photon indices of the two power--law components in the above model 
are assumed to be the same, however their respective normalizations
are allowed to differ. Thus the unabsorbed power--law may represent 
soft X-ray emission which is Thomson scattered into our line of sight, 
while the absorbed power--law represents the intrinsic hard X-ray emission
from an accretion disk corona.
An intrinsic column density of 
$N_{\rm H}=(12.3\pm1.3)\times10^{22}$\,cm$^{-2}$ is required, while the 
derived photon index is still rather flat ($\Gamma=0.78\pm0.13$) compared 
to the typical values in radio--loud AGN (e.g. \citealt{sambruna99}, 
\citealt{reeves00}). 
%The ratio between the normalizations of 
%the unabsorbed to absorbed power-law continua is $f_{\rm scatt}\sim0.14$.

%The resulting data/model ratio to this absorption model is shown 
%in Figure\,2. 

The model is a statistically poor representation of the 
data ($\chi^2/{\rm dof}=421.1/181$, $P=3.2\times10^{-21}$); 
in the soft X-ray band narrow emission--like residuals are present, 
while at high energies
significant spectral curvature is present, as well as emission and absorption 
in the iron K band. The spectral curvature and very hard photon index 
may indicate 
that a more complex 
absorber is present, which was found to be the case in the XMM-Newton spectrum 
of 3C\,445 \citep{sambruna07}. 

Thus instead the spectrum was parameterized by a dual absorber of the form
$F(E) = {\rm wabs} \times ({\rm zwabs}\times{\rm zpcfabs}\times{\rm pow1} + {\rm pow2})$, where 
${\rm zpcfabs}$ is a photoelectric absorber which partially covers our line of sight 
towards the source.
Thus some fraction ($f_{\rm cov}$) of the primary hard X-ray power--law (${\rm pow1}$)
is absorbed by the partial coverer, while the remaining fraction
$(1-f_{\rm cov})$ is absorbed only by the fully covering absorber (${\rm zwabs}$). 
The scattered soft X-ray power-law continuum is absorbed only by the Galactic column, 
as previously described. This dual--absorber provides a good description of the continuum, 
especially above 2 keV where the spectral curvature is no longer present in the data/model 
residuals. The column density of the partial coverer is then 
$N_{\rm H, pcov}=(3.6\pm0.5)\times10^{23}$\,cm$^{-2}$, with a covering fraction of 
$f_{\rm cov}=0.86\pm0.03$, while the fully covering absorber has a lower column 
of $N_{\rm H}=(5.7\pm0.6)\times10^{22}$\,cm$^{-2}$. 
The ratio of the unabsorbed (scattered) to absorbed power--law continuum is 
lower, $f_{\rm scatt}\sim0.01$, consistent with what is seen in 
other absorbed AGN (e.g. Turner et al. 1997).
The power--law photon index 
is now much steeper, with $\Gamma=1.84\pm0.04$. 

Nonetheless the fit statistic is still poor ($\chi^2/{\rm dof}=314.7/179$, $P=1.6\times10^{-9}$), 
with several emission lines apparent below 2 keV (e.g. 
in the O\,\textsc{vii-viii} band) and at Fe K. Indeed evidence for soft X-ray line emission 
has been previously suggested by the XMM-Newton spectra of this 
source (\citealt{sambruna07}; \citealt{grandi07}). 
Below we give a detailed description of these emission lines, 
utilizing the full spectral resolution of the Chandra/LETG. We return in Section 4 
to discuss more physical models for both the absorber and emitter, using the 
photoionization code {\sc xstar} \citep{kallman04}.

\subsection{The Soft X-ray Emission Line Spectrum of 3C 445}

To analyse the emission lines in detail, the LETG spectra were
binned more finely to sample the resolution of the detector, 
at approximately the FWHM spectral resolution (e.g. $\Delta\lambda
=0.05$\AA\ bins). Thus the spectral resolution is $E/\Delta E \sim 500$
(or 600\,km\,s$^{-1}$ FWHM) at 0.5\,keV. 
For the fits, the C-statistic was adopted
\citep{cash79} rather than $\chi^{2}$, as there are  
fewer than 10 counts in some of the resolution bins.

The emission lines were modeled with Gaussian profiles and the best-fit 
partial covering continuum model was adopted from above, allowing 
the continuum and absorption parameters to vary. Table~1
lists the detected lines with their observed and inferred properties,
and their significance as per the C-statistic. Figures~2 and 3
show the portions of the LETG spectrum containing the strongest
lines, with the model overlaid. Overall the fit statistic 
improves considerably upon the addition of the emission lines to 
the continuum model, i.e. from ${\rm C/dof} = 720.2/444$ without 
emission lines (rejected at 
$>99.99$\% confidence) to ${\rm C/dof} = 448.3/425$ upon adding the 
emission lines (rejected at only 85\% confidence).  

Indeed the majority of the individual emission lines in Table~1 and 
are detected at
high confidence, corresponding to $\Delta C>14$, or $>99.9\%$
significance for 2 parameters of interest (note lines 
detected with a lower level of confidence are noted). 
For instance the O\,\textsc{vii} He-$\alpha$ and O\,\textsc{viii} Lyman-$\alpha$ 
lines are detected with $\Delta C > 50$.  
The strongest emission lines correspond to the He-like (He-$\alpha$) 
and H-like (Lyman-$\alpha$) transitions
from O\,\textsc{vii-viii}, Ne\,\textsc{ix-x}, Mg\,\textsc{xi-xii}, and
Si\,\textsc{xiii-xiv}. Fluorescence lines may also be present from 
S\,\textsc{i} K$\alpha$ and Fe\,\textsc{i} K$\alpha$ at 2.3 and 6.4\,keV 
respectively, which may originate from reflection off Compton thick 
matter (see Section 4.1). 
Most of the rest--frame energies of the emission lines are close to their 
expected expected lab values\footnote{Line energies and atomic 
data are adopted from http://physics.nist.gov.} (see Table 1), 
implying that the outflow velocity of the emitting gas is within 
$<1000$\,km\,s$^{-1}$.
%CHECK - ARE THERE ANY EXCEPTIONS HERE.

\subsubsection{Radiative Recombination Emission}

In addition, radiative recombination continua (RRCs) are detected from 
both O\,\textsc{vii} and O\,\textsc{viii} at high ($>99.9$\%) 
significance. 
For instance the emission line detected at 
$700.8\pm1.3$\,eV (17.71\AA) corresponds to a rest frame energy of 
$740.1\pm1.4$\,eV (16.77\AA), which is consistent with the expected 
energy (739.3\,eV) of the O\,\textsc{vii} RRC. 
These have been modeled with emission from recombination edges, of 
variable width dependent on temperature $kT$, 
rather than as Gaussians. Note that both RRCs 
are resolved with a width of $kT\sim3$\,eV (see Table 1), which implies a 
temperature of $\sim3\times10^{4}$\,K for the emitting gas. 
This suggests an origin in a photoionized rather than collisionally 
ionized (thermal) plasma, as the temperature 
would need to be closer to $\sim 10^{7}$\,K to produce 
substantial soft X-ray line emission from collisionally ionized gas.

\subsubsection{The He-Like Line Emission}

The He-like lines all appear to be resolved when modeled with a 
single Gaussian profile (see Table 1). 
Thus the He-like emission may contain a blend of lines from the forbidden (f), 
intercombination (i) and resonance (r) transitions. Indeed the rest energy of the
He-like lines is intermediate between the expected resonance and intercombination 
transitions; e.g. O\,\textsc{vii} at 565\,eV is intermediate between 561\,eV (f) and 
569\,eV (i). Thus the He-like emission was fitted by a blend of 2 separate
emission lines, the results are reported in Table\,2. This provides an acceptable 
parameterization of the spectrum ($C/{\rm dof} = 445.3/423$). The rest energies of the 
O\,\textsc{vii} and Ne\,\textsc{ix} lines are all consistent (within an eV) of the 
expected energies of the forbidden and intercombination transitions. 
%The only exceptions 
%are the possible intercombination lines from Mg\,\textsc{xi} and Si\,\textsc{xiii} 
%which appear somewhat blue-shifted, although a contribution from resonance emission 
%cannot be excluded in this case, which could contribute following photo--excitation 
%\citep{kink02}.
Figures 2 and 3 show the spectrum fitted with this line model.

In all the cases the strengths of the forbidden and intercombination emission is 
approximately equal (see Table\,2), implying a high density plasma, i.e. 
$n_{e}>10^{10}$\,cm$^{-3}$ (\citealt{porquet00}).  The R ratio quantifies the 
ratio of the forbidden to intercombination line strengths; i.e. $R = z/(x+y)$, where $z$ 
is the forbidden line and $x+y$ represents the sum of the intercombination emission. 
In the 3C\,445 spectrum the best constraints on $R$ arises from the O\,\textsc{vii} triplet. 
Allowing the ratio of the forbidden and intercombination lines to vary results in 
a value of $R=0.9^{+1.1}_{-0.3}$ (at 90\% confidence), which implies an electron density 
in the region of $n_{\rm e}=10^{10}-10^{11}$\,cm$^{-3}$ (e.g. Figure 8, \citealt{porquet00}). 
This suggests the emission originates from matter closer in than the Narrow Line 
Region (NLR), a point we discuss further in Section 5. 

%G Ratio

\subsubsection{Line Widths}

The velocity widths of the O\,\textsc{vii} and O\,\textsc{viii} emission lines 
were also determined. The O\,\textsc{viii} Lyman-$\alpha$ line 
is resolved with a width of $\sigma=2.0^{+1.1}_{-0.9}$\,eV (see Table 1), 
corresponding to a FWHM width of $2100^{+1100}_{-950}$\,km\,s$^{-1}$; the fit statistic 
worsens by $\Delta C=12$ if the line 
width is set to zero. Note that the separation of the O\,\textsc{viii} Lyman-$\alpha$ 
doublet has a negligible effect on the line width.
Upon modeling the He-like triplet emission using two separate Gaussian 
lines to represent the 
forbidden and intercombination emission, 
the O\,\textsc{vii} He-$\alpha$ line width was also determined, 
giving $\sigma=2.4^{+2.8}_{-0.8}$\,eV (see Table 1), which 
corresponds to a FWHM width of $3000^{+3400}_{-980}$\,km\,s$^{-1}$. 
Note that the line widths of the forbidden and intercombination lines 
were assumed to be equal to each other in the model.
The fit statistic also 
worsened significantly ($\Delta C=+31$) when the velocity width of the 
O\,\textsc{vii} lines were fixed to zero. 
However the lines from higher Z ions were not resolved, as the LETG 
spectral resolution worsens with increasing energy, however the upper-limits to their
widths are consistent with the values from the Oxygen lines (see Table 1 for the 
H-like ions and Table 2 for the He-like ions).

In order to derive the most accurate determination of the line velocity width, 
we assumed that the widths of the three strongest lines from O\,\textsc{vii}(f), 
O\,\textsc{vii}(i) and O\,\textsc{viii} Lyman-$\alpha$ were identical and tied these values in 
the resulting model. The resolution of the LETG is also at its highest in the O band. 
This yielded a best-fit velocity width of $\sigma=1120^{+430}_{-270}$\,km\,s$^{-1}$ 
(or $\sigma=2.1^{+0.8}_{-0.5}$\,eV at 561\,eV), corresponding to a FWHM width of 
$v_{\rm FWHM}=2600^{+1000}_{-600}$\,km\,s$^{-1}$.
A contour plot showing the 
measurement of the FWHM velocity width of the Oxygen lines is shown in Figure 4, 
which shows that a velocity width of zero is 
excluded at $>99.99$\% confidence (with $\Delta C = +42$).

\section{Photoionization Modeling}

As discussed above, the presence of strong radiative recombination continua may 
suggest that the soft X-ray line emission originates from a photoionized rather than 
collisionally ionized plasma \citep{kink02}. 
To test this the spectrum was fitted with 
a collisional model, such as by the \textsc{mekal} \citep{kaastra93} or \textsc{apec} 
\citep{smith01} codes. We used the same continuum parameterization as before, except 
the ionized emission lines modeled by Gaussians are replaced by emission from a 
collisionally ionized 
\textsc{apec} model. A single temperature collisional model 
of temperature $kT=0.24\pm0.04$\,keV does not provide an acceptable fit to the data 
($C/{\rm dof}=610.5/456$) and the model is rejected at $>99.99$\% confidence. 
Figure 5 (upper panel) shows the spectrum fitted with the \textsc{apec} model 
in the Oxygen band, the model fails to account for the O\,\textsc{vii} 
triplet (the resonance line is the strongest predicted line, which is not present in 
the actual data), while the model does not fit the recombination 
emission.
We also investigated whether a multiple temperature collisionally ionized model or a model 
with non--Solar abundances further improved the fit, but this was not the case.
Thus the data appear to exclude a collisionally ionized plasma for the origin 
of the soft X-ray emission lines.

\subsection{The Soft X-ray Emitter}

%Thus instead we used a grid of emission models calculated by 
%the photoionization code \textsc{xstar} v2.1 \citep{kallman82, kallman04}
%to derive the parameters of the emitter, assuming the baseline
%continuum described above. Solar abundances
%are assumed throughout \citep{gs98}. A turbulent velocity of 
%$\sigma=500$\,km\,s$^{-1}$ has been used for the emission model\footnote{We have 
%chosen a grid whereby the turbulence velocity best matches the observed velocity 
%width from the Gaussian line fits.} 
%We assumed an initial column density of $N_{\rm H}=1.8\times^{23}$\,cm$^{-2}$ 
%for the emitter, which is equal to the absorber column density found 
%from the photoionized absorption model in Section 4.2. Note that the 
%emitter column density
%cannot be directly determined from the spectrum, 
%as the $N_{\rm H}$ and total covering factor of the emitter are largely degenerate 
%upon each other. 
%Likely values for both the column and covering of the emitter 
%will be discussed further in Section 5.1.2 

Thus instead we used a grid of emission models calculated by 
the photoionization code \textsc{xstar} v2.1 \citep{kallman82, kallman04}
to derive the parameters of the emitter, assuming the baseline
continuum described above. Solar abundances
are assumed throughout \citep{gs98}. A turbulent velocity of 
500\,km\,s$^{-1}$ has been used for the emission model.\footnote{We have 
chosen a grid whereby the turbulence velocity best matches the observed velocity 
width from the Gaussian line fits.}
We assumed an initial column density of $N_{\rm H}=10^{22}$\,cm$^{-2}$ 
for the emitter as 
this cannot be fitted directly to the data, unlike for an absorption model, 
as the $N_{\rm H}$ and total covering factor of the emitter are largely degenerate 
upon each other. Likely values for both the column and covering of the emitter 
will be discussed further in Section 5.1.2 

The overall model fitted to the data is in the same 
form as the partial covering model described in Section 3.1, i.e. 
$F(E) = {\rm wabs} \times ({\rm zwabs}\times{\rm zpcfabs}\times{\rm pow1} + {\rm pow2} + 
{\rm xstar_{em}})$, 
where ${\rm xstar_{em}}$ represents the photoionized emission, which is absorbed only 
by the Galactic line of sight column.
The \textsc{xstar} model does not include the fluorescent emission 
from S K$\alpha$ and Fe K$\alpha$, which have been modeled separately 
with Gaussian profiles.
This provides a significantly better 
fit to the data than the collisional model; for comparison Figure 
5 (lower) shows that the \textsc{xstar} model fits the Oxygen emission lines well, 
especially around the O\,\textsc{vii} triplet and RRC, in contrast to the collisional 
model. The overall fit statistic is improved to $C/{\rm dof}=529.1/450$, which is 
rejected only at the 90\% confidence level. The best fit parameters of this 
\textsc{xstar} model are listed in Table\,3. 

The emitter was fitted by 2 zones of gas of ionization parameter of 
$\log \xi=1.8^{+0.1}_{-0.3}$ and $\log \xi=3.0\pm0.4$ respectively\footnote{The ionization 
parameter is defined here as $\xi=L_{\rm ion}/n_{\rm e}R^{2}$, where $L_{\rm ion}$ is the 
ionizing luminosity from 1--1000\,Rydberg, $n_{\rm e}$ is the electron density and $R$ 
is the radial distance to the gas. The units of $\xi$ are erg\,cm\,s$^{-1}$.}.
Note that a two zone model gave only a 
slightly better fit (by $\Delta C=9$ for 2 parameters) than a single zone model, 
thus a second higher ionization emission 
zone is only required at the $\sim99$\% confidence level. 
There is also some tentative 
evidence for super--Solar abundances of Mg and Si (all other abundances are consistent 
with Solar), although their values in Table\,2 are not well determined.

Note no outflow velocity is required for the photoionized emitter, indeed a slight redshift 
is found with $v_{\rm out}=+150^{+240}_{-210}$\,km\,s$^{-1}$, although the data are 
consistent with zero velocity shift (compared to systemic) for the emitter. 
Thus the 90\% confidence limit to the emitter outflow velocity is very 
tightly constrained to within 60\,km\,s$^{-1}$ of the systemic velocity of 3C\,445.
In comparison the emitter outflow velocity determined from the XMM-Newton/RGS 
data was $v_{\rm out}=-430^{+220}_{-160}$\,km\,s$^{-1}$ \citep{grandi07}.

Finally instead of modeling the Fe K$\alpha$ line with a Gaussian, a 
model consisting of Compton reflection off an optically--thick photoionized 
slab of gas was tested. 
The \textsc{reflionx}\footnote{see http://heasarc.gsfc.nasa.gov/docs/xanadu/xspec/models/reflion.html} model was used (\citealt{ross99}; \citealt{ross05}), 
assuming Solar abundances, 
with the resulting fit found to be equally good compared to the fit with Gaussian 
fluorescent lines. The reflector was found to be low ionization, with an upper-limit of 
$\log \xi<1.7$. This is not surprising given that the rest energy of the iron K$\alpha$ 
emission (see Table 1) is close to the expected value for neutral or lowly ionized iron 
(i.e. Fe\,\textsc{i-xvii}). Note that no velocity broadening 
is required for the reflection spectrum, which is consistent with the upper--limit 
to the iron K$\alpha$ velocity width of $\sigma<145$\,eV (or $\sigma<6800$\,km\,s$^{-1}$). 
Thus the reflection spectrum could be consistent with an origin in either the outer 
accretion disk, 
or a pc-scale Compton-thick torus. The properties of the reflection component 
are discussed in more detail in a paper describing the Suzaku and Swift spectrum 
and the hard X-ray emission above 10\,keV \citep{braito10}.

\subsection{The X-ray Absorber}

The neutral partial covering model provides a good 
phenomenological description of the absorber in 3C\,445, whereby part of the 
primary X-ray emission is heavily absorbed 
(by a column of gas of $N_{\rm H}>10^{23}$\,cm$^{-2}$) 
along the line of sight. However this 
may also approximate a scenario whereby the absorber is partially 
ionized and thus can be partially transparent to continuum X-rays at soft X-ray energies.
Indeed the majority of Seyfert 1 spectra show such a ``warm'' absorber (e.g. 
\citealt{reynolds97}; \citealt{blustin05}; \citealt{mckernan07}). Here we are more
likely to be viewing the nucleus of 3C\,445 at higher inclination angles of about 
$60-70$ degrees, 
given the likely radio orientation of the system \citep{erac98, sambruna07}.
Thus in 3C\,445 we may be viewing the X-ray source 
directly down an accretion 
disk wind \citep{gallagher07} or perhaps through the outer edge of the 
putative pc-scale molecular torus \citep{urry95}.

Initially, as a simple parameterization, 
the iron K band absorption was fitted by a simple photoelectric edge model. 
The energy of the edge was found to be $E=7.35\pm0.09$\,keV in the rest--frame 
of 3C\,445, with an optical depth of $\tau=1.0\pm0.2$. Thus crudely 
parameterized in this way, the edge energy is significantly greater than the 
K-shell threshold energy for neutral iron at 7.11\,keV, which indicates that 
the iron K absorber towards 3C\,445 may be at least partially ionized.

Thus the neutral partially covering absorber (the model \textsc{zpcfabs} in \textsc{xspec}), 
as well as the neutral fully covering absorber, 
were instead replaced by a single photoionized absorber, in the form of a multiplicative 
grid of absorption models calculated by the \textsc{xstar} v2.1 code. 
The \textsc{xstar} model includes the detailed treatment of the iron K-shell 
opacity as described by \citet{kallman04}.
Solar abundances were assumed and the absorber was assumed to fully cover the 
line of sight to the source. Thus the overall spectral model was in the 
mathematical form:- 
$F(E) = {\rm wabs} \times [{\rm xstar_{abs}}\times({\rm pow1} + {\rm reflionx}) 
+ {\rm pow2} + {\rm xstar_{em}}]$, where ${\rm xstar_{abs}}$ represents the 
multiplicative \textsc{xstar} photoionized absorber, ${\rm reflionx}$ is the 
Compton reflection component and ${\rm pow1}$, ${\rm pow2}$, ${\rm xstar_{em}}$ and 
${\rm wabs}$ are the direct and scattered power-laws, photoionized emitter and 
neutral Galactic absorber respectively, as described previously.

The photoionized absorption model provides an excellent fit to the LETG 
spectrum, the overall fit statistic is $C/{\rm dof}=498.4/450$, 
which is formally acceptable. In comparison the fit statistic for the 
neutral partial covering model is somewhat worse, with $\Delta C=+30$ for 
the same degrees of freedom ($C/{\rm dof}=529.1/450$). Nonetheless the 
partial covering model cannot be rejected with a high degree of certainty, 
as the overall fit statistic is only excluded at 90\% confidence. 
The difference between the two absorber models is shown in Figure 6, it can be 
seen the partially ionized absorber provides a better fit to the absorption 
in the iron K band and in particular the position and depth of the iron K absorption profile, 
although otherwise the fit to the data appears similar.

%with an improvement of $\Delta C=32$ compared to the partial covering 
%absorber. The overall fit statistic is $C/{\rm dof}=499.3/449$, 
%which is formally acceptable. 
%Thus this is considered to be the final 
%best-fit model to the 3C\,445 spectrum, 

The parameters of the photoionized absorber model 
are listed in Table\,3.
A column density of $N_{\rm H}=(1.85^{+0.09}_{-0.11}) \times 10^{23}$\,cm$^{-2}$ 
is found for the 
photoionized absorber, which is moderately ionized, with $\log \xi=1.42^{+0.20}_{-0.12}$. 
At this ionization, iron ions in the range Fe\,\textsc{xv-xviii} are likely 
to dominate the spectrum \citep{kallman04}, although other ions 
may also be present. At this ionization state, the absorber consists of a 
blend of resonance lines from the above ions, similar to the model spectra 
that are shown in Figure 13 of \citet{kallman04}, resulting in a broad 
absorption trough as observed in the LETG above 7 keV. 
Note however that calorimeter based 
resolution in the iron K band would be required to resolve the absorption line 
structure.

Interestingly, the best fit model appears to indicate that the absorber 
could be outflowing compared to the systemic 
velocity of 3C\,445, with $v_{\rm out}=-(0.034\pm0.002)c$ 
(or $v_{\rm out}=-10200\pm600$\,km\,s$^{-1}$).
Note a solution with zero outflow velocity is excluded 
at $>99.9$\% confidence and the fit statistic is correspondingly worse by 
$\Delta C = 22.3$ in this case. The apparent blueshift of the absorber 
is driven by the requirement to fit the absorption profile above 7\,keV in the 
iron K band. 

Figure 7 shows the overall fit statistic (C-statistic) 
for the absorption model, plotted against 
the redshift of the photoionized absorber, obtained from stepping 
through the absorber redshift in small increments of $\Delta z = 10^{-3}$. Note 
all the other parameters of the absorber (e.g. $N_{\rm H}$ and $\log \xi$) and the
continuum were also allowed to vary at each increment.
Thus an absorber at a redshift of $z=0.0562$ would require no net velocity shift 
compared to the host galaxy of 3C\,445; however an absorber redshift of 
$z=0.0562$ appears to be excluded at $>99.99$\% confidence from the fit statistic.
Indeed the best fit absorption model to the Chandra spectrum has a redshift of 
$z=0.022\pm 0.002$. 
An intervening absorption system at an intermediate 
redshift of $z=0.022$ would appear unlikely, as this would require the whole 
X-ray spectrum of 3C\,445 to be absorbed, rather than just the primary power-law,  
which cannot be the case as the soft X-ray line emission is not absorbed 
by the $N_{\rm H}\sim10^{23}$\,cm$^{-2}$ column of gas. Furthermore no intervening 
absorption systems are known at this redshift towards 3C\,445. Thus the most 
likely scenario is that the photoionized absorber in 3C\,445 is outflowing, 
with a net blue-shift of approximately $-10000$\,km\,s$^{-1}$ with respect to the rest--frame 
of 3C\,445. 

A further more highly ionized zone of absorbing gas 
is not statistically required in the LETG data, 
however the Suzaku data may indicate the presence of 
such absorption at iron K, in the form of resonance absorption from 
Fe\,\textsc{xxv} or Fe\,\textsc{xxvi} \citep{braito10}, with a similar outflow 
velocity to that required above. 
Note that the Galactic ($z=0$) absorption column is in excess of the expected 
value from neutral H\,\textsc{i} measurements (e.g. \citealt{dickey90}), which 
might indicate additional neutral absorption associated with the host galaxy of 
3C\,445. 

Finally the presence of a scattered power--law component is required in the 
model at high confidence ($\Delta C = 56$), 
which is not absorbed by the high column density photoionized absorber. 
The fraction of the scattered to direct power--law emission is $\sim2$\%. Thus about 
2\% of the direct absorbed power-law is scattered into our line of sight 
by free electrons in a highly ionized plasma. As we discuss below such 
gas may be 
associated with the photoionized emission region detected in the Chandra spectrum.
%Alternatively the photoionized gas may instead partially cover the central 
%X-ray source, but with a high covering fraction of 98\%.

\section{Discussion}

\subsection{The Properties of the Soft X-ray Emitting Gas}

The high resolution Chandra LETG observation of the BLRG 3C\,445 has revealed a 
complex X-ray spectrum of this obscured AGN, with both emission and 
absorption present from layers of photoionized gas. 
Initially we turn our attention to the properties of the soft X-ray line 
emitting gas. While the presence of the soft X-ray line emission 
has been detected in a previous observation with XMM-Newton (\citealt{sambruna07}; 
\citealt{grandi07}), here for the first time we detect and resolve multiple 
emission lines, utilizing the high spectral 
resolution of the Chandra LETG. 
Specifically the Chandra LETG observation has resolved emission 
lines from highly ionized gas in the soft X-ray spectrum, primarily from 
the He and H-like transitions of O, Ne, Mg and Si, corresponding to 
the most abundant elements with K-shell emission lines over 
the 0.5--2.0\,keV band. The direct detection of radiative recombination 
emission from O\,\textsc{vii} and O\,\textsc{viii} also shows that the most 
likely origin of the soft X-ray line emission is from photoionized emission; 
indeed a model including 
emission from a collisionally ionized gas is unable to produce an 
acceptable fit to the spectrum.

\subsubsection{The Distance and Density of the Gas}

The emission from the He-like triplets of O\,\textsc{vii}, 
Ne\,\textsc{ix}, Mg\,\textsc{xi} and Si\,\textsc{xiii} appear to be resolved 
into their respective forbidden and intercombination lines.
Indeed all of the He-like lines are well modeled in this manner, e.g. O\,\textsc{vii} 
for example, as discussed in Sections 3 and 4, while the line energies are in agreement 
with the expected rest--frame energies of the forbidden and intercombination 
transitions. The fact that significant 
intercombination emission is detected suggests the density of the plasma is high. 

Indeed the ratio $R$ of the forbidden to intercombination emission is close to $R\sim1$ 
(see Section 3.2.2). 
This sets a lower-limit on the density of the He-like emitting gas of 
$n_{\rm e}>10^{10}$\,cm$^{-3}$ (see \citealt{porquet00}, Figure 8). Thus an 
upper--limit of the radial distance to the emitter from central X-ray source can be 
estimated from the definition of the ionization parameter of the emitting gas, 
i.e. $R^{2} = L_{\rm ion} / \xi n_{\rm e}$. 
Here $L_{\rm ion}$ is the ionizing luminosity 
from $1-1000$\,Rydberg, which from the best--fit continuum parameters in Table\,3, gives 
$L_{\rm ion} = 3\times 10^{44}$\,erg\,s$^{-1}$ for 3C\,445. The ionization parameter 
of the He-like emitting gas is $\log \xi=1.8$\,erg\,cm\,s$^{-1}$, as measured from the 
\textsc{xstar} emission model (Section 4.1, Table 3), while $n_{\rm e}>10^{10}$\,cm$^{-3}$ 
as described above. Thus the radial distance to the emitter derived from 
this method is $R<2\times 10^{16}$\,cm (or $<0.01$\,pc). For an estimated 
black hole mass of 3C\,445 of $M_{\rm BH} \sim 2\times10^{8}$\Msun\ \citep{bettoni03, marchesini04}, 
this corresponds to 
a radius of $\sim1000 R_{\rm g}$\footnote{where $R_{\rm g}=GM/c^2$ is the gravitational 
radius}. 
%Note that at this radii, the emitting gas is gravitationally bound to the AGN.

Alternatively the location of the emitting gas can be estimated from the measured line 
widths of the O\,\textsc{vii-viii} emission.
The best fit width of the three strongest 
emission lines (namely O\,\textsc{vii} f and i and O\,\textsc{viii} Lyman-$\alpha$)
is $v_{FWHM}=2600^{+1000}_{-600}$\,km\,s$^{-1}$ (section 3.2.3).
Thus assuming 
Keplerian motion, $R \sim GM_{\rm BH} / v^{2}$, where here we define the velocity width 
as $v=\frac{\sqrt{3}}{2} v_{\rm FWHM}$.
Thus by this estimate $R \sim 5\times 10^{17}$\,cm (or $R \sim 0.1$\,pc).
%At this radial distance then the gas density is 
%$n_{\rm e} = L_{\rm ion} / \xi R^{2}$ and thus $n_{\rm e} \sim 10^{9}$\,cm$^{-3}$; 
%hence the density and radius are similar to the above estimate.

%Note that the H-like emission lines are unblended and have a lower intrinsic 
%velocity width than the He-like lines, 
%e.g. O\,\textsc{viii} Lyman--$\alpha$ has an intrinsic width 
%of $v_{\rm FWHM}=2100$\,km\,s$^{-1}$ (see Table 1) and may originate from gas at a 
%greater radial distance, in this case $R \sim 10^{18}$\,cm  
%and may have a lower density (i.e. $n_{e} \sim 10^{6}$\,cm$^{-3}$.)

\subsubsection{The Covering Factor}

We now consider the possible covering factor and column density of the 
emitting gas. From the \textsc{xstar} code\footnote{see http://heasarc.gsfc.nasa.gov/docs/software/xstar.docs/html/node94.html}, the normalization ($k$)
of the line emission component (which is proportional to the flux received by the 
observer) is related to the global covering fraction of a quasi-spherical 
shell of gas by:-

\begin{equation}
k = f_{\rm cov} L_{\rm ion}(10^{38}) / D_{\rm kpc}^{2}
\end{equation}

where $f_{\rm cov}$ is the global covering factor of the emitter ($f_{\rm cov}=1$ 
for a spherical shell covering $4\pi$\,steradians), $L_{\rm ion}(10^{38})$ 
is the ionizing luminosity of the source in units of $10^{38}$\,erg\,s$^{-1}$ 
and $D_{\rm kpc}$ is the distance to 3C\,445 in units of kpc. 
Thus the units of $k$ are $10^{38}$\,erg\,s$^{-1}$\,kpc$^{-2}$.
Note the normalization of the \textsc{xstar} emission component (or equivalently 
its flux) will be lower if the covering fraction of the gas is lower.

The Hubble flow distance to 3C\,445 ($z=0.0562$) is $D=2.37\times10^{5}$\,kpc
assuming $H_0=71$\,km\,s$^{-1}$\,Mpc$^{-1}$. 
From the spectral fits in Section 4 (summarized in Table\, 3), the 
normalizations of the two \textsc{xstar} emission components 
are $k_{\rm low}=2.4\times10^{-6}$ for the lower ionization gas and 
$k_{\rm high}=1.2\times10^{-5}$ for the higher ionization gas. Thus 
for an ionizing luminosity of $3\times10^{44}$\,erg\,s$^{-1}$, the 
covering fractions of the low and high ionization emission regions 
are 0.045 and 0.22 respectively. 

Note that this calculation is for a column density of $N_{\rm H}=10^{22}$\,cm$^{-2}$ 
for the emitter, which has been assumed in the \textsc{xstar} model. 
For the emitter, the column density and covering factor are largely degenerate 
upon each other, in other words if a lower column is assumed then the 
covering fraction of the gas will need to be correspondingly higher 
to compensate, in order to reproduce the same observed 
emission line spectrum. Nonetheless the 
covering fraction cannot be $f_{\rm cov}>1$, hence a lower limit to 
the column density of $N_{\rm H}>2.2\times10^{21}$\,cm$^{-2}$ is derived 
for the high ionization gas. 

For a plausible upper limit to the column density of the emitting gas, 
one may consider the 
fraction of the observed X-ray power-law continuum that is electron scattered 
into our line of sight. From the best-fit model listed in Table\,3, the 
ratio of the scattered to directly observed continuum is 0.02. Note that 
the presence of such a scattered component is required by the data at a high 
confidence level. Thus the fraction of scattered photons is equivalent to 
$1-\exp({-\sigma_{\rm T} N_{\rm H}}) = 0.02$, where 
$\sigma_{\rm T}=6.65\times10^{-25}$\,cm$^{2}$ is the Thomson 
cross-section. Hence the column required to reproduce the 
scattered power-law emission is $N_{\rm H}=3\times10^{22}$\,cm$^{-2}$. 
Note that this electron 
scattering zone may be associated with the highest ionization gas in the emission 
line region (which may be more extended than the lower ionization region), 
which scatters the primary X-ray continuum back into our line of sight.
Such gas is indeed envisaged in AGN Unification scenarios \citep{antonucci93} 
and may be responsible for the broad, permitted lines seen in polarized light 
in the optical spectra of type 2 AGN.

%In the \textsc{xstar} model we have initially assumed that the 
%emitter column density is equal to the absorber column density  
%of $N_{\rm H} = 1.8\times10^{23}$\,cm$^{-2}$.
%From the spectral fits in Section 4 (summarized in Table\, 3), the 
%normalizations of the two \textsc{xstar} emission components 
%are $k_{\rm low}=2.8\times10^{-7}$ for the lower ionization gas and 
%$k_{\rm high}=1.2\times10^{-6}$ for the higher ionization gas. Thus 
%for an ionizing luminosity of $3\times10^{44}$\,erg\,s$^{-1}$, the 
%covering fractions of the low and high ionization emission regions 
%are $5\times10^{-3}$ (0.5\%) and 0.023 (2.3\%) respectively. 

%Note that the emitter column density need not be equal to the absorber 
%column density, as the emitter may not be along the direct line of sight.
%For the emitter, the column density and covering factor are largely degenerate 
%upon each other, in other words if a lower column is assumed then the 
%covering fraction of the gas will need to be correspondingly higher 
%to compensate, in order to reproduce the same observed 
%emission line spectrum. Nonetheless the 
%covering fraction cannot be $f_{\rm cov}>1$, hence a lower limit to 
%the emitter column density of $N_{\rm H}>2.2\times10^{21}$\,cm$^{-2}$ is derived 
%for the high ionization gas. 
 
The approximate mass of the emitting gas can also be estimated. The total mass 
of the emitter is equal to $M = 1.23f_{\rm cov} \times 4\pi R^{2} N_{\rm H} m_{\rm p}$, 
where the factor $\times 1.23$ arises from the Solar composition of the gas. 
We adopt the radius of the emitting gas derived in Section 5.1.1 from the He-like triplets, of 
$R \sim 0.01$\,pc. 
From the possible range of column densities discussed above 
of $N_{\rm H}=(2\times10^{21} - 3\times10^{22})$\,cm$^{-2}$, 
corresponding to covering fractions of 
between $f_{\rm cov} = 1.0 - 0.08$ respectively, 
then the emitter mass is only $M \sim 0.03$\Msun.
However the mass will be considerably higher for larger $R$, 
i.e. if the larger radius of $R \sim 0.1$\,pc is adopted from the Keplerian width of the 
Oxygen lines (see Section 5.1.1), then the mass will be of the order $M \sim 3$\Msun.

\subsubsection{Emission from an X-ray Broad Line Region?}

Thus the possible radius derived for emitting gas of $R \sim 0.01 - 0.1$\,pc,
appears to be well inside the expected radii for the Narrow Line Region 
(i.e. pc to kpc scales) and also within the expected size-scale of 
putative pc-scale molecular torus. 
Furthermore the likely density of this gas, of $n_{\rm e} \sim 10^{10}$\,cm$^{-3}$, 
appears much higher than what one 
would associate with typical NLR densities (e.g. $n_{\rm e} \sim 10^{3}$\,cm$^{-3}$, 
\citealt{koski78}).
 
Instead the range of radii and densities calculated above appear to be 
consistent with what is typically expected in the AGN Broad Line Region of Seyfert 
1 galaxies and quasars (e.g. \citealt{wandel99}; \citealt{kaspi00}; \citealt{peterson04}). Indeed the velocity width measured here of $\sim 2600$\,km\,s$^{-1}$ (FWHM) is similar to 
the measured H$\alpha$ FWHM width for 3C\,445 of $6400$\,km\,s$^{-1}$
(Eracleous \& Halpern 1994), or the H$\beta$ FWHM of 3000\,km\,s$^{-1}$ 
(Osterbrock, Koski \& Phillips 1976). 
%Furthermore the estimated covering factor and mass of the X-ray emitting gas, 
%appears to be consistent 
%with the predicted covering fraction and mass of 
%the optical BLR clouds (e.g. \citealt{netzer93}).

Recently, broadened soft X-ray lines were detected by \citet{longinotti08} 
during an XMM-Newton/RGS observation of the Seyfert 1 Mrk 335 during a low flux state. 
The O\,\textsc{viii} Lyman-$\alpha$ line was resolved with a FWHM width 
of $2200\pm750$\,km\,s$^{-1}$, consistent with the H$\beta$ width of this Seyfert\,1 
\citep{boroson92}. From photoionization arguments, \citet{longinotti08} placed a 
limit on the radial distance to the soft X-ray emitting gas of $<0.06$\,pc, consistent with 
a BLR origin. A similar claim has been made by \citet{blustin09}, who detected 
several broadened soft X-ray emission lines with a typical FWHM of $\sim 5000$\,km\,s$^{-1}$, 
from a deep XMM-Newton/RGS observation of the narrow lined Seyfert 1, 1H\,0707-495. 
Indeed there have also been several past claims of broadened soft X-ray emission lines from 
grating observations of Seyfert 1s, with line widths which appear to be 
consistent with a BLR origin 
(e.g. NGC 4051, \citealt{ogle04}; NGC 5548, \citealt{steenbrugge05}; Mrk 509, \citealt{smith07}; 
Mrk 279, \citealt{cost07}). Thus we may be observing the same X-ray BLR emission in
the BLRG 3C 445.

\subsection{Absorption from an Accretion Disk Wind in 3C 445?}

As has also been found from previous observations 
(\citealt{sambruna07}; \citealt{grandi07}), the primary X-ray continuum 
from 3C\,445 
is highly absorbed, with a column density exceeding $10^{23}$\,cm$^{-2}$. 
Indeed the X-ray column observed towards 3C\,445 
far exceeds the expected column based on the 
extinction in the optical band towards of this AGN, of the order
$E_{B-V} \sim 1$ \citep{rudy82}. 
Although the properties of the
absorbing gas was unclear from previous shorter (and lower spectral resolution) 
observations, the Chandra LETG shows the 
absorption can be well modeled with a moderately ionized outflow, 
with an outflow velocity of the order $\sim 10000$\,km\,s$^{-1}$. As we 
will discuss here, it appears more plausible that this absorbing gas is associated with a 
disk wind on smaller (sub--parsec) scales rather than with a putative 
molecular torus.

\subsubsection{The Likely Location of the Wind}

We first consider the location of the absorber. For a homogeneous radial
wind, the observed column density along the line of sight is equal to:-

\begin{equation}
N_{\rm H} = \int_{\rm R_{in}}^{\rm R_{out}} n_{\rm e}(R) dR = 
\int_{\rm R_{in}}^{\rm R_{out}} \frac{L_{\rm ion}}{\xi R^{2}} dR
=\frac{L_{\rm ion}}{\xi}\left(\frac{1}{R_{\rm in}} - \frac{1}{R_{\rm out}}\right)
\end{equation}

where $R_{\rm in}$ and $R_{\rm out}$ are the inner and outer radii 
along the line of sight through the wind. In the case where we are looking 
directly down a homogeneous wind towards an inner radius $R_{\rm in}$, 
then $R_{\rm out}=\infty$ and re-arranging gives
$R_{\rm in} = L_{\rm ion} / \xi N_{\rm H}$. The best fit \textsc{xstar} 
model of the absorber (Table\,3) gives $N_{\rm H}=2\times10^{23}$\,cm$^{-2}$ 
and $\log \xi = 1.4$, thus $R_{\rm in} = 5\times 10^{19}$\,cm (or $\sim 10$\,pc). 
At this radius the density of the absorbing matter is 
$n_{\rm e} \sim 10^{4}$\,cm$^{-3}$.

Thus a radius of $\sim 10$\,pc 
is consistent with a wind launched from the region of the molecular 
torus. However this can be considered an upper-limit, 
if we are only viewing across the wind, or if the wind is sufficiently clumpy, as 
we will argue in Section 5.3 from considering the wind energetics. In this case 
we can define the wind thickness as $\Delta R = R_{\rm out} - R_{\rm in}$ 
and assuming $\Delta R / R <<1$, then re-arranging equation\,2 gives:-

\begin{equation}
R_{\rm in} = \frac{L_{\rm ion}}{\xi N_{\rm H}}\left(\frac{\Delta R}{R}\right)
\end{equation} 

So if $\Delta R / R <<1$ then the wind 
can originate on much more compact scales.  
Indeed if the location of the X-ray absorber is on parsec scales, then this 
presents a problem for both the soft X-ray emission line gas and 
for the emission from the optical BLR.
Thus the location of the absorber cannot be outside of the soft X-ray line emission 
region, as otherwise the emission lines would be completely obscured. 
This is certainly not the case from the LETG spectrum, i.e. the soft X-ray 
emission is absorbed by a much lower column density of 
$N_{\rm H} = 1.5\times10^{21}$\,cm$^{-2}$ (see Table 3). 
Furthermore the emission from the optical broad line 
region in 3C\,445 cannot be strongly obscured by the X-ray absorbing gas, 
given its classification as a BLRG and the observation 
of broad permitted lines such as H$\alpha$ or H$\beta$ in non-polarized light 
\citep{erac94}.
Thus the observational evidence suggests a more compact size-scale 
for the X-ray absorber in 3C\,445 than the pc-scale torus.

For a scenario whereby the absorber originates from a wind launched 
off the accretion disk by radiation pressure, we 
can calculate a lower bound on the launch radius from the escape velocity.
%Here we assume the wind is already full accelerated at the radius it is 
%observed. 
The escape radius is $R_{\rm esc} = 2c^{2} R_{\rm g} / v_{\rm out}^{2}$, 
where $R_{\rm g}=3\times10^{13}$\,cm (for
$M_{\rm BH}=2\times10^{8}$\Msun) and $v_{\rm out}=10^{4}$\,km\,s$^{-1}$. Thus for 
the outflow in 3C\,445, $R_{\rm esc} \sim 3\times10^{16}$\,cm (or $\sim 0.01$\,pc). 
This would appear to be a more plausible radius at which the wind is viewed, 
as while the central X-ray continuum would be absorbed, the sight--line towards 
the soft X-ray emission need not be obscured. At this radius the 
density would be higher, $n_{\rm e}\sim10^{10}$\,cm$^{-3}$, while the gas
will be substantially clumped, i.e. $\Delta R/R \sim 10^{-3}$. 

Indeed such a scenario could also be consistent with the absorbing clumps or 
clouds partially covering the line of sight to the central X-ray emitter. 
In this scenario, in order for the 
clouds to only partially obscure the primary X-ray continuum, they would likely have to 
be only a few gravitational radii in size, i.e. of the order $\sim 10^{14}$\,cm. 
If the column density through an individual cloud is $N_{\rm H} \sim 10^{23}$\,cm$^{-2}$, 
this would require a density of $n \sim 10^{9}$\,cm$^{-3}$. Given an observed ionization 
parameter of $\log \xi=1.4$, this then implies a radial distance to the clouds of 
$\sim 10^{17}$\,cm, i.e. consistent with a sub-parsec scale location.

\subsection{The Wind Energetics}

For a quasi--spherical radial outflow, the mass outflow rate is given by:-

\begin{equation}
\dot{M}_{\rm out} = 1.23 \times 4\pi b R^{2} n_{\rm e} m_{\rm p} v_{\rm out} 
= 1.23 \times 4\pi b \left(\frac{L_{\rm ion}}{\xi}\right) m_{\rm p} v_{\rm out}
\end{equation} 

where for the absorber in 3C\,445, $L_{\rm ion} / \xi = 10^{43}$\,cm$^{-1}$, while $b$ 
is a geometrical factor, where $b=1$ for a homogeneous spherical outflow, but $b<1$ 
for a wind covering a fraction of $4\pi$ steradian or if the matter is clumped.
Thus for the absorber in 3C\,445, then 
$\dot{M}_{\rm out} = 2.5b \times 10^{29}$\,g\,s$^{-1}$ or 
$\dot{M}_{\rm out} = 4000b$\Msun\,yr$^{-1}$. This seems implausibly 
high for $b=1$. 

For comparison we can estimate the likely accretion 
rate needed to power the bolometric (radiative) output of 3C\,445. 
The (unabsorbed) 2--10\,keV X-ray luminosity of 3C\,445 
is $L_{2-10}=1\times10^{44}$\,erg\,s$^{-1}$ and assuming the 2-10\,keV luminosity 
is approximately 5\% of the bolometric output (e.g. \citealt{elvis94}), 
then $L_{\rm bol}=2\times10^{45}$\,erg\,s$^{-1}$, also in agreement with the 
estimate of Marchesini et al. (2004). For an accretion efficiency of 
$\eta=0.05$, then the accretion rate is then 
$\dot{M}_{\rm acc} = L_{\rm bol} / \eta c^{2} \sim 1$\Msun\,yr$^{-1}$ for 
3C\,445. Thus a homogeneous wind requires 
$\dot{M}_{\rm out} >> \dot{M}_{\rm acc}$ for 3C\,445, which would rather 
rapidly exhaust the supply of gas towards the central AGN.

Instead we consider the outward transfer of momentum into the wind, via 
radiation pressure. Thus the outward momentum rate is
$\dot{p} = \dot{M}_{\rm out} v_{\rm out} = L_{\rm bol} / c \sim 10^{35}$\,g\,cm\,s$^{-2}$.
Thus for the measured outflow velocity of $v_{\rm out} =10^{9}$\,cm\,s$^{-1}$, then the 
mass outflow rate is $\dot{M}_{\rm out} \sim 10^{26}$\,g\,s$^{-1}$ (or $\sim 1$\Msun\,yr$^{-1}$).
This appears more physically realistic with $\dot{M}_{\rm out} \sim \dot{M}_{\rm acc}$ and 
requires a clumping factor of $b\sim 10^{-3}$, which may be the case if the absorbing 
clouds are located on sub--parsec scales as discussed above. Furthermore we note that 
the Suzaku spectrum of 3C\,445 \citep{braito10} may require a very high 
ionization zone ($\log \xi \sim 4$) of absorbing gas 
in addition to the moderately ionized absorber discussed here. 
Indeed both the low and high ionization absorption may co--exist at the same radii 
in a multi--phase disk--wind, with the low ionization absorber present as higher density 
clouds confined within the lower density and more 
uniform high ionization outflow (see Figure 8).

Finally we estimate the total mechanical output of the wind in 3C\,445. 
The kinetic power is simply 
$\dot{E} = \dot{M}_{\rm out} v_{\rm out}^{2}/2 \sim 10^{47}b$\,erg\,s$^{-1}$, 
which for $b \sim 10^{-3}$, as discussed above, implies $\dot{E} 
\sim 10^{44}$\,erg\,s$^{-1}$, which is approximately 10\% of $L_{\rm bol}$.
Note that the presence of a high ionization outflow of $\log \xi \sim 4$, 
as implied by the Suzaku data, also suggests a similar mechanical output, but 
without the requirement that the absorbing gas is clumped.

\subsection{The Overall Geometry of 3C 445}

Given the likely high inclination of 3C\,445 compared to the radio-jet 
axis, of the order $60-70$\deg\ \citep{leahy97, erac98}, 
it is plausible that we are viewing the central engine of 3C\,445 at a relatively 
side-on orientation. The situation may be similar to that outlined in Figure 8. 
Thus we may be viewing through an equatorial 
disk-wind in 3C 445, on scales of approximately $10^{16}-10^{17}$\,cm from the
central black hole. In this toy model for 3C\,445, the soft X-ray emitting clouds may be 
lifted above the plane of the accretion disk and could 
be associated with the optical BLR emission, as discussed earlier.
Any highly ionized gas that is present would also serve to Thomson scatter the 
primary absorbed X-ray continuum back into our line of sight, as is also observed in the 
Chandra spectrum.

The lower ionization absorbing gas viewed here may well be clumped, 
within a much higher ionization, but lower density medium, co-existing at a similar 
radius to the high ionization gas. 
If such high column density ($\sim 10^{23}$\,cm$^{-2}$) but low ionization
absorbing gas were located on much larger scales, 
e.g. with a parsec scale torus, it would be 
difficult to reconcile the high column density with 
the much lower absorbing column (of $N_{\rm H} \sim 10^{21}$\,cm$^{-2}$), 
towards the soft X-ray line emitting gas, if the latter is 
coincident with sub-parsec BLR scales. Instead 
the much lower column density gas which absorbs 
the soft X-ray emission spectrum might instead be associated with gas on the scale of the 
host galaxy and would be consistent with the amount of reddening observed towards 
3C\,445 \citep{rudy82}.

Finally if the high column absorbing medium is clumped, it might be possible to 
observe short-timescale $N_{\rm H}$ variability, due to the passage of clouds along 
our line of sight.
For instance such aborption variability 
is observed along the line of sight to some Seyfert galaxies, the most notable example 
being the intermediate type Seyfert, NGC 1365 \citep{risaliti09, maiolino10}. 
However no such column density variations have been apparent so far towards 3C\,445, either 
on short timescales within the 200\,ks Chandra observations, or on longer timescales,  
as the total column density is also consistent with previous observations, e.g. with XMM-Newton 
(\citealt{sambruna07}; \citealt{grandi07}). 
However this may simply be explained if the number of absorbing clouds is large, 
which would average out any variations.

\section{Conclusions}

We have reported upon a deep 200\,ks Chandra LETG spectrum of the absorbed 
BLRG, 3C\,445, which displays a complex X-ray spectrum. The high resolution 
Chandra spectrum revealed a wealth of soft X-ray emission lines from a 
photoionized plasma, primarily 
from the He and H-like transitions of O, Ne, Mg and Si. The 
O\,\textsc{vii} and O\,\textsc{viii} lines are resolved, 
with a typical FWHM width of $\sim 2600$\,km\,s$^{-1}$, while the ratio of forbidden 
to intercombination emission in the He-like triplets indicate a high electron 
density, of $n_{\rm e}>10^{10}$\,cm$^{-3}$ \citep{porquet00}.
Thus the X-ray lines appear to 
be consistent with an origin in the optical BLR in 3C\,445, located on sub-parsec scales.

The Chandra spectrum of 3C\,445 is also highly absorbed and can be modeled by 
either partially covering or by partially ionized absorbing gas. The 
high column density gas may be associated with an equatorial accretion disk 
wind in 3C\,445, observed at high inclinations with respect to the radio jet axis. 
Future high resolution observations of 3C\,445, with calorimeter resolution in the 
iron K band, e.g. with Astro-H, will potentially resolve a host of absorption 
lines associated with the high column density absorber, enabling us to probe the 
kinematics of any outflowing wind to a high level of accuracy.

\acknowledgements

This research has made use of data obtained from the High Energy
Astrophysics Science Archive Research Center (HEASARC), provided by
NASA's Goddard Space Flight Center.
%and of the NASA/IPAC Extragalactic
%Database (NED) which is operated by the Jet Propulsion Laboratory,
%California Institute of Technology, under contract with the National
%Aeronautics and Space Administration. 
R.M.S. acknowledges support from
NASA through the \suzaku\ and \chandra\ programs. 
%{\bf Rita - do you have a Chandra grant to acknowledge?}.
%M.E. thanks the NSF for support via grant AST-0807993.
We would also like to thank 
Tahir Yaqoob for assistance with the Chandra data analysis.

\clearpage

\begin{figure*}
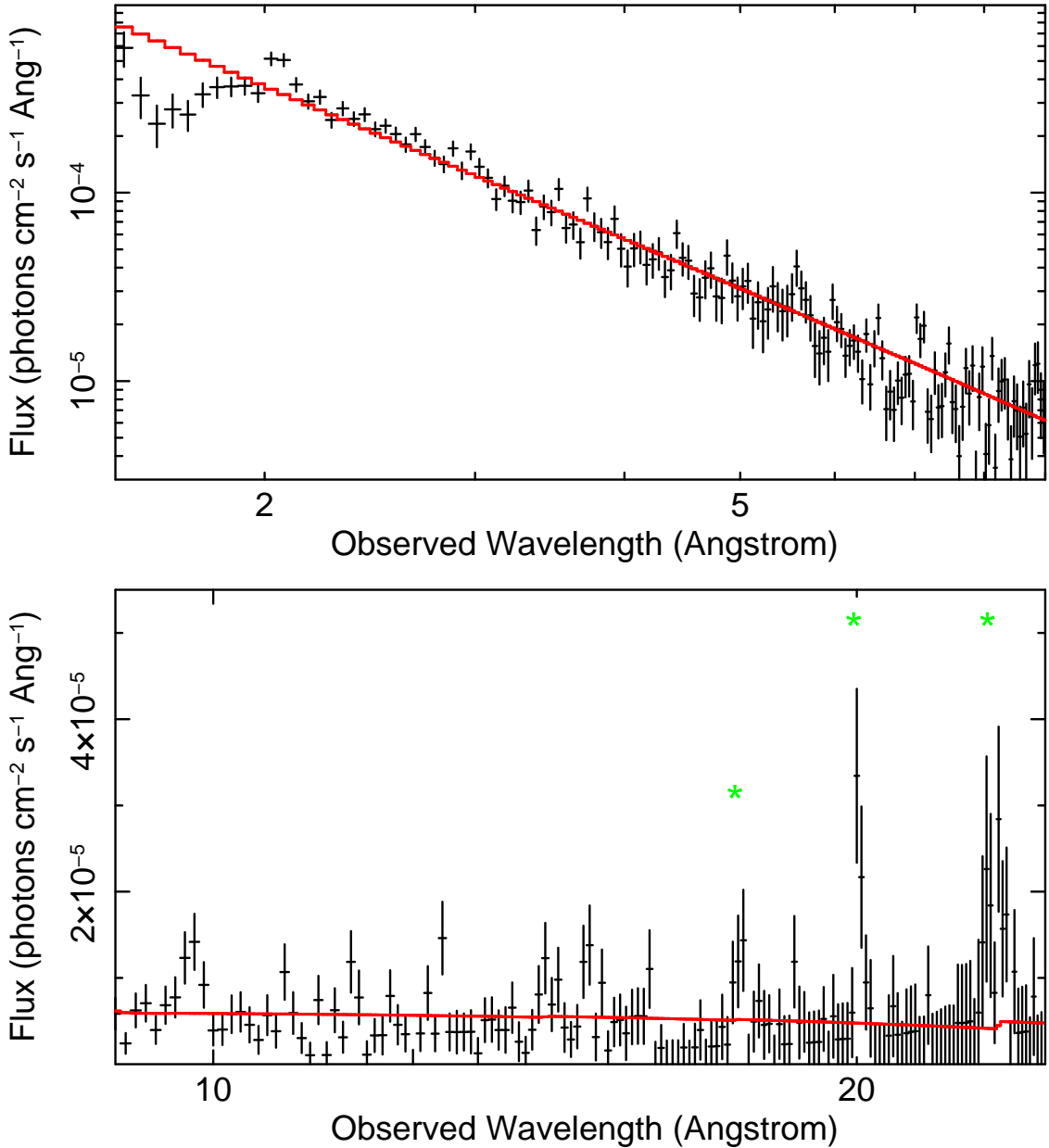

\begin{center}
\rotatebox{-90}{\includegraphics[height=15cm]{f1a.eps}}
\rotatebox{-90}{\includegraphics[height=15cm]{f1b.eps}}
\end{center}
\caption{The Chandra LETG spectrum of 3C\,445, showing the hard X-ray (top) and 
soft X-ray regions (lower) against observed--frame wavelength. The fluxed 
spectrum (black crosses) is plotted against a broken power-law continuum (solid line) 
for reference. The spectrum shortwards of 9\AA\ ($>1.4$\,keV) is very hard, 
rising with energy (decreasing wavelength)
with a photon index of $\Gamma=-0.64$, indicative of a highly absorbed 
continuum. Note the emission and absorption present at iron K. 
In the soft X-ray band several emission lines appear to be present. The star symbols 
denote the expected positions of (from right to left) the 
O\,\textsc{vii} He-$\alpha$ (forbidden or intercombination), O\,\textsc{viii} 
Lyman-$\alpha$ and the O\,\textsc{vii} radiative recombination emission.} 
\end{figure*}

%\begin{figure*}
%\begin{center}
%\rotatebox{-90}{\includegraphics[height=15cm]{f2a.eps}}
%\rotatebox{-90}{\includegraphics[height=15cm]{f2b.eps}}
%\end{center}
%\caption{The Chandra LETG spectrum of 3C\,445, plotted as a
%data/model ratio to an absorbed power-law continuum 
%model of the form
%$\rm{wabs} \times (\rm{pow} + \rm{zwabs} \times \rm{pow}$),
%as 
%described in the text. The fit is clearly unacceptable and several emission lines are clearly present in the residuals between 0.5--3.0\,keV, as well as at 
%iron K.} 
%\end{figure*}
 
\begin{figure*}
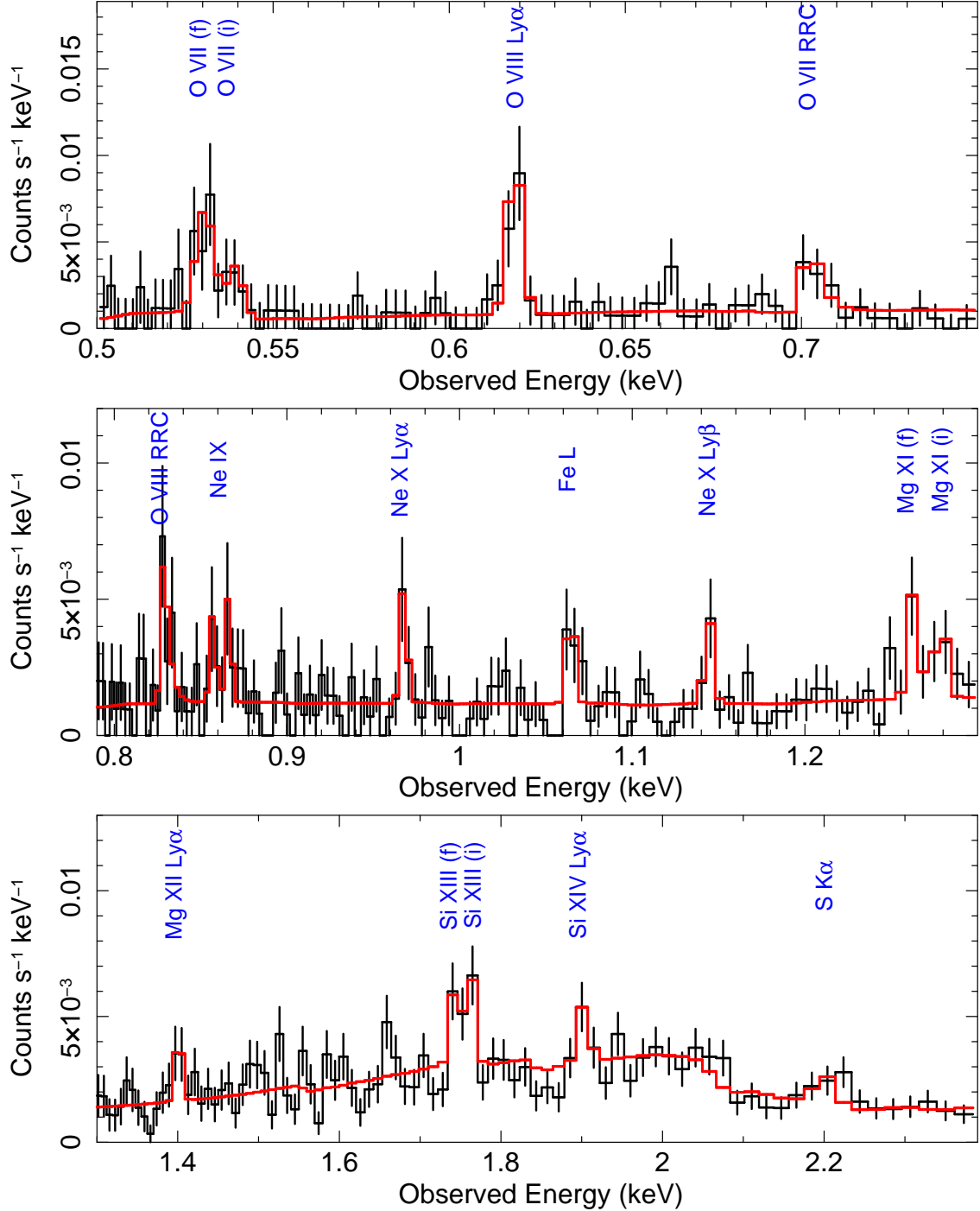

\begin{center}
\rotatebox{-90}{\includegraphics[height=15cm]{f2a.eps}}
\rotatebox{-90}{\includegraphics[height=15cm]{f2b.eps}}
\rotatebox{-90}{\includegraphics[height=15cm]{f2c.eps}}
\end{center}
\caption{Chandra LETG spectra of 3C 445 showing several soft X-ray emission lines from O, Ne, Mg, Si and 
Fe L. The solid line represents the best-fit absorbed continuum model with Gaussian emission lines, 
as described in Section 3.2 and Tables 1 and 2.} 
\end{figure*}

\begin{figure*}
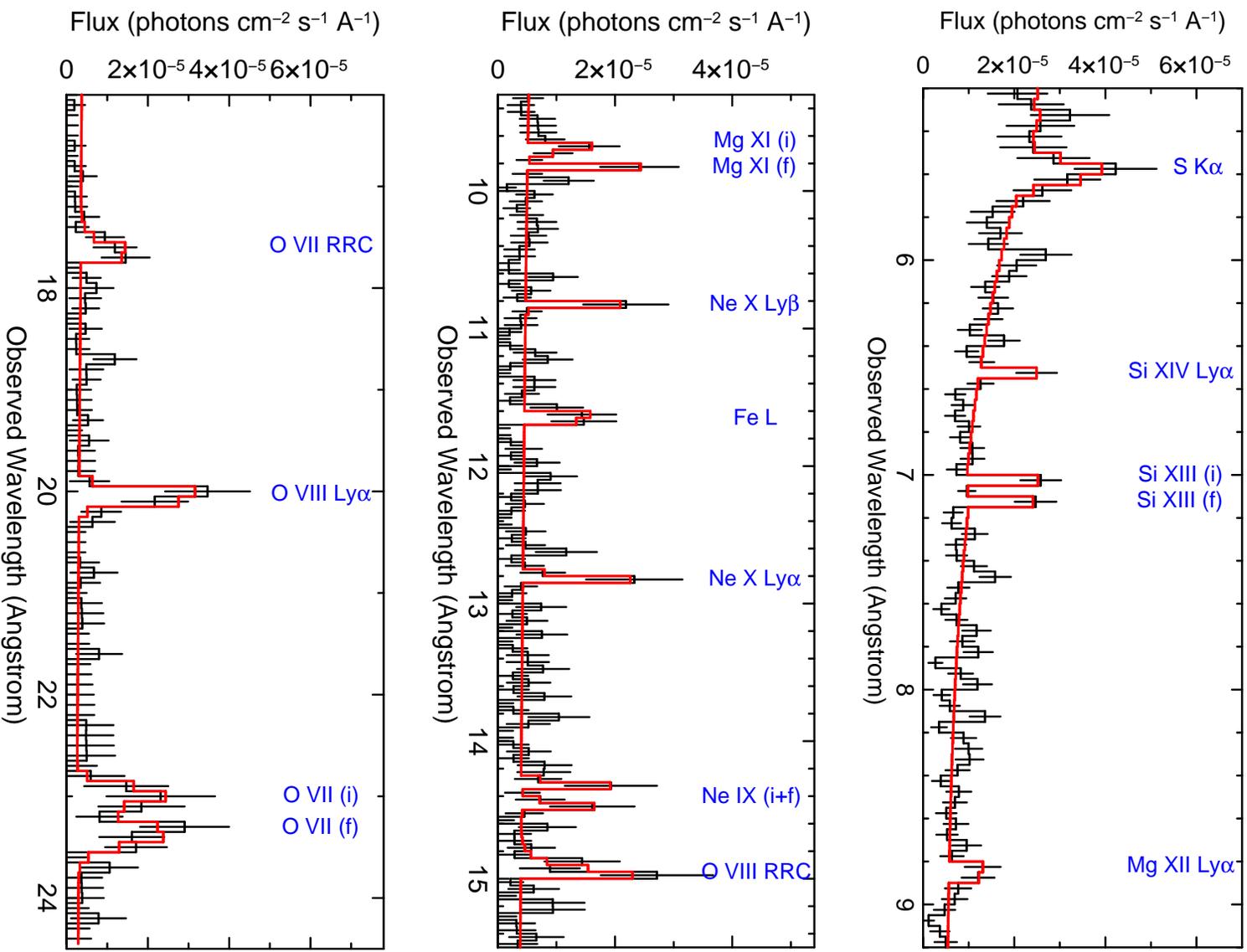

\begin{center}
\rotatebox{-90}{\includegraphics[height=15cm]{f3a.eps}}
\rotatebox{-90}{\includegraphics[height=15cm]{f3b.eps}}
\rotatebox{-90}{\includegraphics[height=15cm]{f3c.eps}}
\end{center}
\caption{Fluxed LETG spectrum, plotted against observed--frame 
wavelength (in Angstroms). The solid line shows the same line model as 
plotted in Figure 2.} 
\end{figure*}

\begin{figure}
\begin{center}
\rotatebox{-90}{
\epsscale{0.7}
\plotone{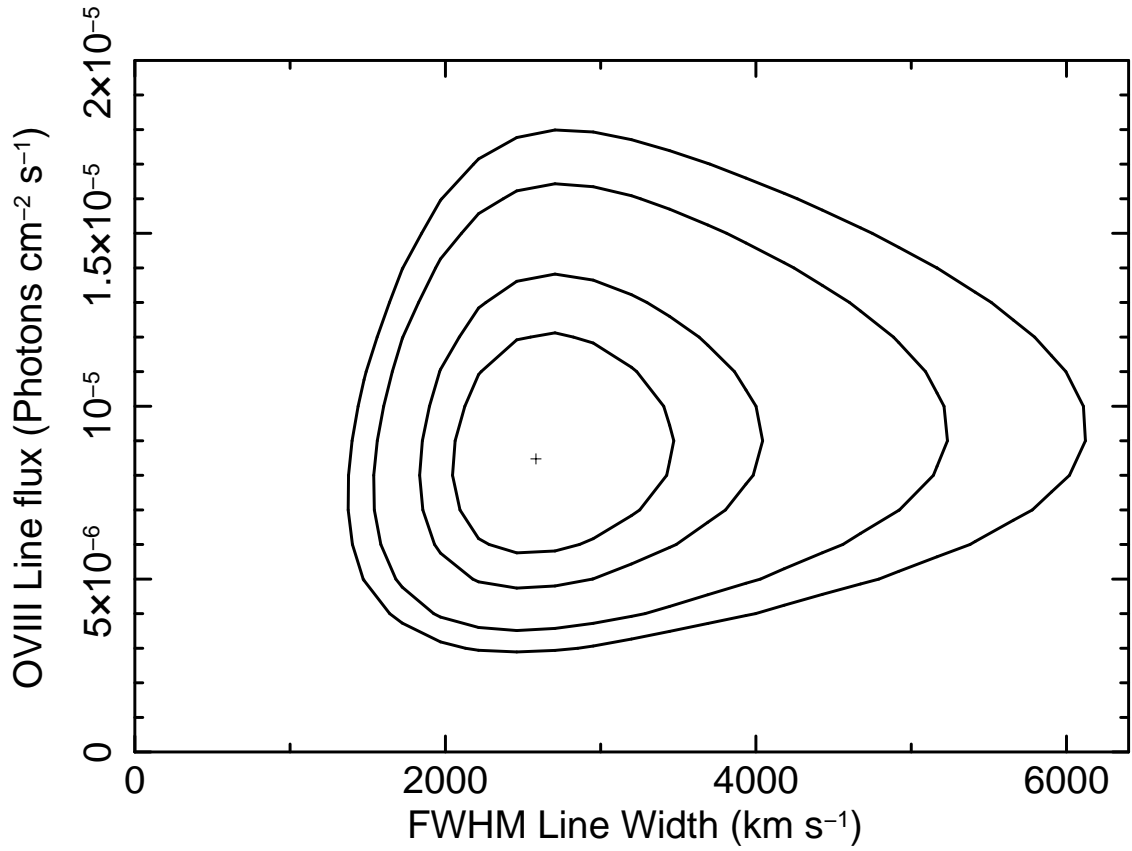}}
\caption{FWHM velocity width, fitted jointly to the 
O\,\textsc{vii} forbidden/intercombination and O\,\textsc{viii}
Lyman-$\alpha$ lines, plotted against O\,\textsc{viii}
Lyman-$\alpha$ line flux. The contours represent the 68\%, 90\%, 
99\% and 99.9\% confidence levels for 2 parameters of interest.}
\end{center}
\end{figure}

\begin{figure*}
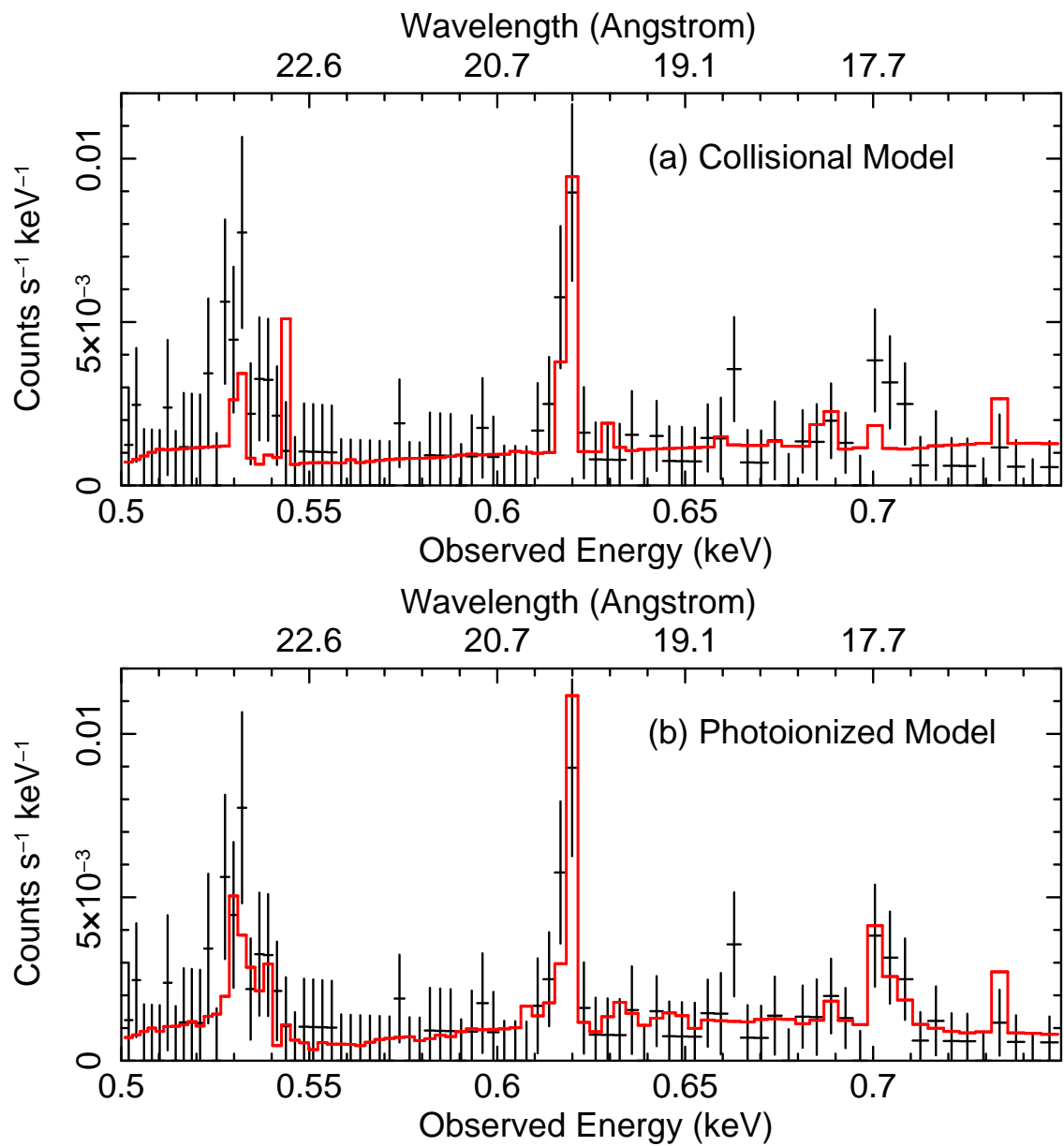

\begin{center}
\rotatebox{-90}{\includegraphics[height=15cm]{f5a.eps}}
\rotatebox{-90}{\includegraphics[height=15cm]{f5b.eps}}
\end{center}
\caption{Comparison between (a) the collisionally ionized \textsc{apec} 
emission model and (b) the photoionized \textsc{xstar} emission model in the 
Oxygen band, as discussed in Section 4.} 
\end{figure*}

\begin{figure*}
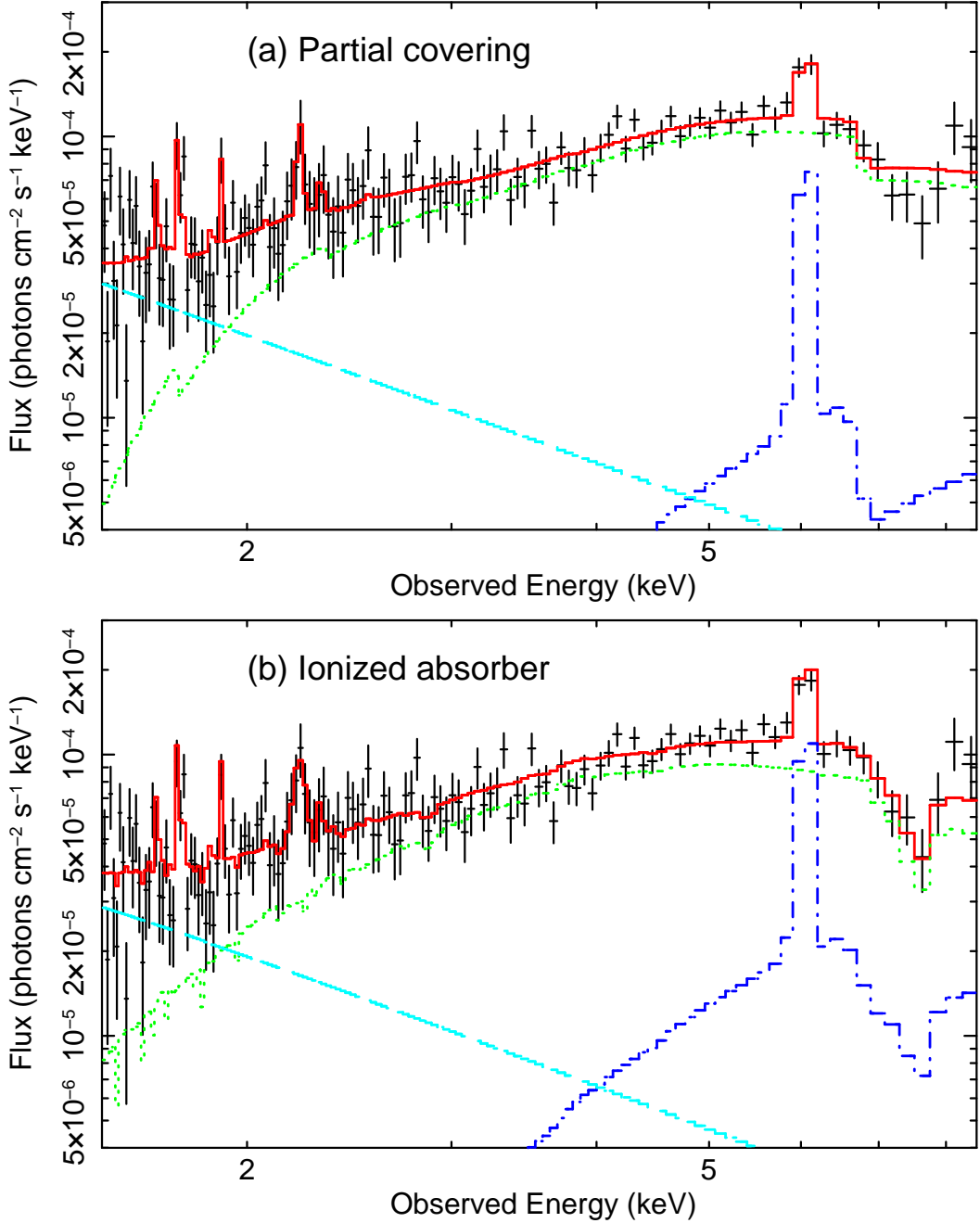

\begin{center}
\rotatebox{-90}{\includegraphics[height=14cm]{f6a.eps}}
\rotatebox{-90}{\includegraphics[height=14cm]{f6b.eps}}
\end{center}
\caption{Chandra LETG spectrum plotted in the iron K band. Panel (a) 
shows the spectrum fitted with a neutral partial covering absorption model, 
as described in Section 4.2. Panel (b) shows the spectrum fitted with 
a partially ionized absorber, also described in Section 4.2. The solid (red) line 
shows the total emergent spectrum (including emission lines), 
the dotted (green) line the absorbed 
power-law, the dot-dashed (blue) line the reflection component and the dashed 
(cyan) line the unabsorbed scattered power-law.
Note that the 
partially ionized absorber gives a better fit in the iron K band.} 
\end{figure*}

\begin{figure}
\begin{center}
\rotatebox{-90}{
\epsscale{0.7}
\plotone{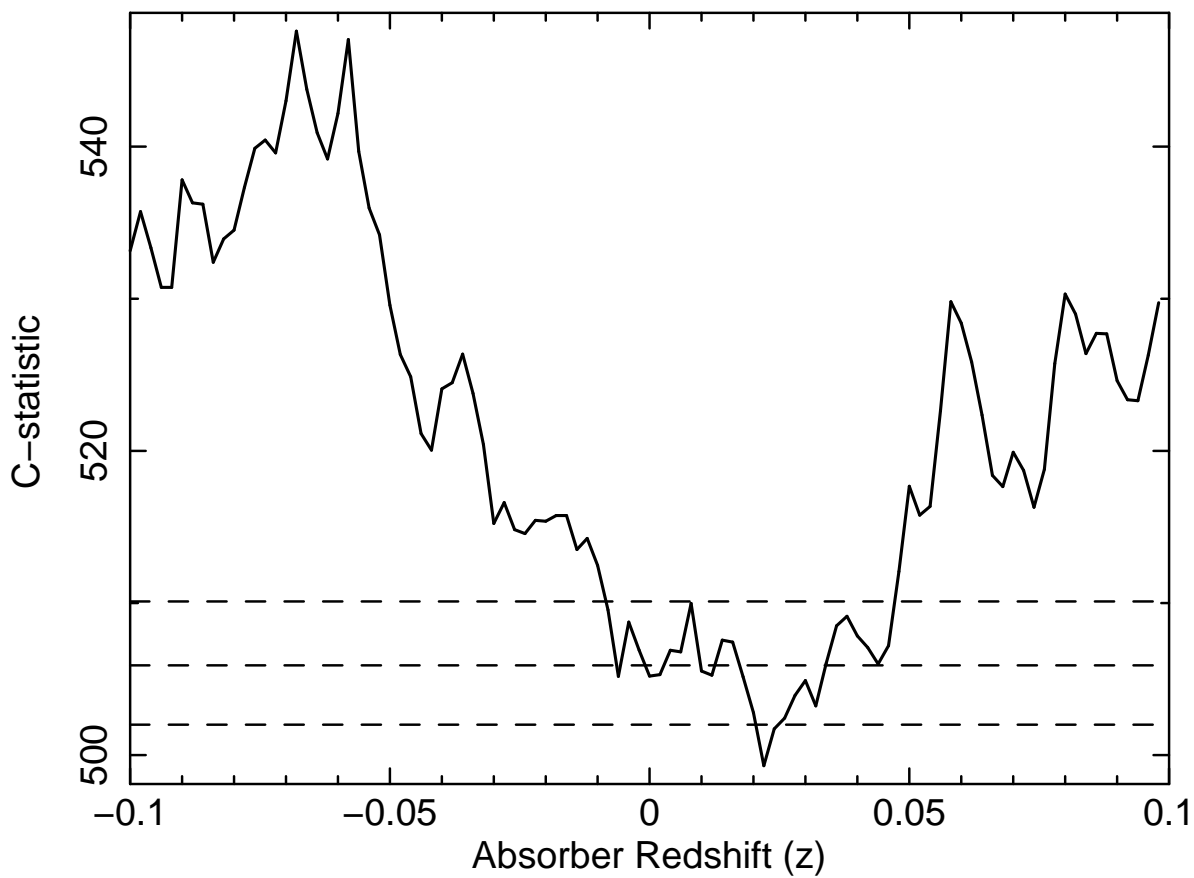}}
\caption{Overall fit statistic (C-statistic) plotted against absorber 
redshift (solid line). 
An absorber coincident with the redshift of 3C\,445 at $z=0.056$ appears 
to be ruled out at 
$>99.9$\% confidence level. Note the dashed horizontal lines represent (from bottom to top) 
the 90\%, 99\% and 99.9\% confidence levels, for 1 interesting parameter. 
The best-fit absorber redshift 
of $z=0.022\pm0.002$ implies that the absorber has a net blueshift of $-10000$\,km\,s$^{-1}$, 
in the rest frame of 3C\,445.}
\end{center}
\end{figure}

\begin{figure}
\begin{center}
\rotatebox{0}{
\epsscale{0.75}
\plotone{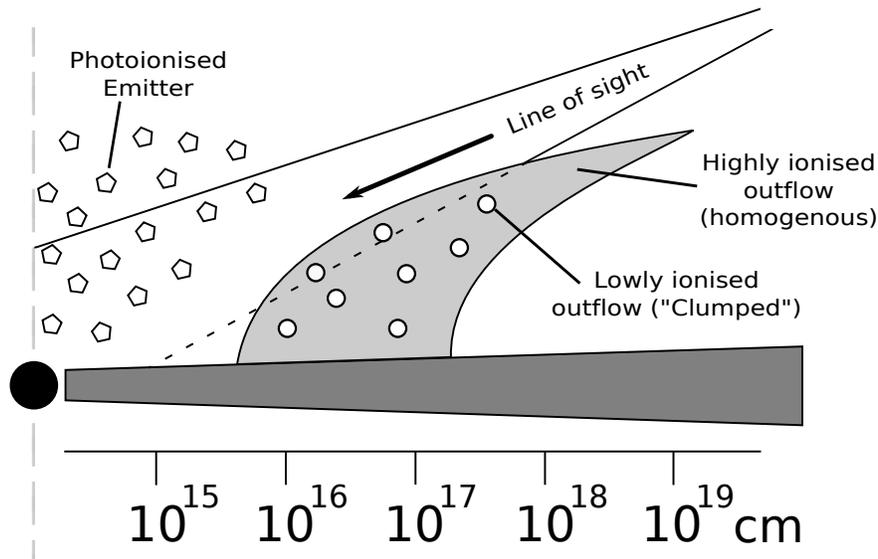}}
\caption{Schematic diagram showing the possible inner geometry  
of 3C\,445. The line of sight towards the observer is represented by the solid and dashed 
black lines, while the grey dashed vertical line represents the direction of the radio-jet axis. 
In 3C\,445 the line of sight towards 
the direct X-ray emission from the innermost region (black dashed line) is obscured by the outflowing 
matter, which may be in the form of a clumpy accretion disk wind. 
The observer has an unobscured 
view of the photoionized emission line clouds, which are responsible for the soft X-ray line emission. 
This highly ionized gas can also scatter continuum X-rays into the line of sight.}
\end{center}
\end{figure}

\clearpage
\begin{deluxetable}{lccccccl}
\tabletypesize{\small}
\tablecaption{Summary of LETG emission line parameters.}
\tablewidth{0pt}
\tablehead{
\colhead{$E_{\rm rest}$\,(eV)} & \colhead{$\lambda_{\rm rest}$\,(\AA)} & 
\colhead{Line Flux$^{a}$} & \colhead{$\sigma$ or $kT$\,(eV)$^{b}$} 
& \colhead{EW (eV)} & \colhead{$\Delta C^{c}$} & \colhead{Line ID} & \colhead{$E_{\rm lab}^{d}$\,(eV)}
}
\startdata

$564.6\pm2.6$ & $21.98\pm0.10$ & $18.8^{+7.6}_{-6.2}$  & $6.3^{+2.8}_{-1.8}$ & $105^{+42}_{-35}$  
& 55.0  & O VII He$\alpha$ & 561.1(f) \\ [0.5ex]

$653.4\pm1.0$ & $18.99\pm0.03$ & $8.3^{+4.0}_{-3.0}$   & $2.0^{+1.2}_{-0.9}$ & $62^{+29}_{-22}$  & 54.4 & O VIII Ly$\alpha$ & 653.7 \\ [0.5ex]

$740.2\pm1.4$ & $16.76\pm0.03$ & $3.2^{+2.4}_{-1.7}$ & $3.5^{+3.6}_{-1.9}$ & $31^{+22}_{-15}$  & 18.8  & 
O VII RRC & 739.3 \\ [0.5ex]

$873.8\pm1.4$ & $14.20\pm0.02$ & $2.2^{+1.5}_{-1.1}$   & $3.0^{+4.0}_{-2.0}$ & $21^{+12}_{-10}$  & 19.6  
& O VIII RRC & 871.4 \\ [0.5ex]

$913\pm5$ & $13.59\pm0.07$ & $2.6^{+1.6}_{-1.3}$  & $7.2^{+4.0}_{-2.0}$ 
& $25^{+15}_{-12} $ & 14.6 & Ne IX He$\alpha$ & 905.1(f) \\ [0.5ex]

$1022^{+1}_{-2}$ & $12.14^{+0.01}_{-0.02}$ & $1.3^{+1.0}_{-0.8}$ & $<4.0$ & $19^{+15}_{-12}$  & 14.2 & Ne X Ly$\alpha$ & 1022 \\ [0.5ex]

$1124 \pm 2$ & $11.04\pm0.02$ & $1.2^{+0.9}_{-0.7}$   & ---  & $20^{+14}_{-12}$ & 11.8$^{e}$  
& Fe XXIII & 1125 \\ [0.5ex]

$1209^{+3}_{-2}$ & $10.26\pm^{+0.03}_{-0.02}$ & 
$0.9^{+0.8}_{-0.6}$   & --- & $18^{+16}_{-12}$  & 10.7$^{e}$ 
& Ne X Ly$\beta$ & 1211 \\ [0.5ex]
 
$1341\pm7$ & $9.25\pm0.05$ & $2.6^{+1.2}_{-1.0}$   & $14.5^{+9.0}_{-5.0}$  
& $59^{+27}_{-23}$  & 26.1 & Mg XI He$\alpha$ & 1331(f) \\ [0.5ex]

$1480\pm3$ & $8.38\pm0.02$ & $0.8^{+0.6}_{-0.5}$ & $<21$ & $21^{+16}_{-13}$  & 10.2$^{e}$ 
& Mg XII Ly$\alpha$ & 1472  \\ [0.5ex]

$1853\pm6$ & $6.97\pm0.02$ & $1.4^{+0.6}_{-0.5}$   & $10.5^{+5.5}_{-4.0}$  & $32^{+14}_{-11}$ & 23.7  & Si XIII He$\alpha$ & 1839(f) \\ [0.5ex]

$2010^{+6}_{-10}$ & $6.17^{+0.02}_{-0.03}$  & $0.7^{+0.5}_{-0.4}$   & $<13$  & $13^{+16}_{-9}$ & 9.5$^{e}$ 
& Si XIV Ly$\alpha$ & 2006 \\ [0.5ex] 

$2343^{+16}_{-9}$ & $5.30^{+0.04}_{-0.02}$ & $2.5^{+1.5}_{-1.3}$   & $<38$  & $43^{+25}_{-22}$  & 11.5$^{e}$ & S I K$\alpha$ & 2307\\ [0.5ex]

$6364^{+40}_{-43}$ & $1.95\pm0.01$ & $22.9^{+6.6}_{-6.1}$  & $<145$ & $191^{+55}_{-51}$ & 55.6  & Fe I K$\alpha$                \\ [1.0ex]
\hline
\enddata

\tablenotetext{a}{Measured Flux in units of $10^{-6}$ photons cm$^{\rm{-2}}$ s$^{\rm{-1}}$} 
\tablenotetext{b}{$1\sigma$ Width of Gaussian line or temperature of radiative recombination continua (in eV). 
If only an upper-limit to the width is determined, the width has been fixed at $\sigma=1$\,eV 
in the model.}
\tablenotetext{c}{Improvement in $C$ to fit after adding emission line component}
\tablenotetext{d}{Expected line energy of transition taken from http://physics.nist.gov.}
\tablenotetext{e}{Line detection significance at lower than 99.9\% confidence.}
\end{deluxetable}

\clearpage

\begin{deluxetable}{lccccccc}
\tabletypesize{\small}
\tablecaption{Summary of He-like Triplets}
\tablewidth{0pt}
\tablehead{
\colhead{$E_{\rm rest}$ (eV)} & \colhead{$\lambda_{\rm rest}$\,(\AA)} &
\colhead{Line Flux$^{a}$} & \colhead{$\sigma$\,(eV)$^{b}$} & \colhead{EW\,(eV)} & \colhead{$\Delta C^{c}$} & \colhead{Line ID$^{d}$}
& \colhead{$E_{\rm lab}$\,(eV)}
}

\startdata

$560.7\pm1.4$ & $22.13\pm0.06$ & $8.6^{+4.7}_{-3.7}$ & $2.4^{+2.8}_{-0.8}$ & $27^{+23}_{-12}$  & 35.2     & O VII(f) & 561.1\\ [0.5ex]

$569.5\pm1.6$ & $21.79\pm0.06$ & $9.2^{+7.2}_{-4.1}$ & -- & $21^{+20}_{-11}$  & 16.0     & O VII(i) & 568.7\\ [0.5ex]

$905.4 \pm 2.3$ & $13.70\pm0.03$ & $1.0^{+1.0}_{-0.7}$ & $<5$ & $8^{+10}_{-6} $   & 6.5$^{e}$ & Ne IX(f) & 905.1\\ [0.5ex]

$914.8\pm2.0$ & $13.56\pm0.03$& $1.2^{+1.0}_{-0.7}$ & -- & $13^{+21}_{-10}$  & 9.0$^{e}$ & Ne IX(i) & 914.8\\ [0.5ex]

$1334^{+2}_{-3}$ & $9.30^{+0.01}_{-0.02}$ & $1.1^{+0.7}_{-0.5}$ & $<21$ & $19^{+12}_{-9}$   & 16.4   & Mg XI(f) & 1331 \\ [0.5ex]

$1351^{+6}_{-3}$ & $9.18^{+0.04}_{-0.02}$ & $0.8^{+0.7}_{-0.5}$ & -- & $15^{+13}_{-9}$   & 9.7 & Mg XI(i or r) & 1343 or 1352\\ [0.5ex]

$1842^{+4}_{-5}$ & $6.74\pm0.02$  & $0.8^{+0.5}_{-0.4}$ & $<6$  & $16^{+10}_{-8}$   & 12.0 & Si XIII(f) & 1839\\ [0.5ex]

$1861 \pm 3$ & $6.67\pm0.01$ & $1.1^{+0.5}_{-0.4}$ & -- & $25^{+11}_{-9}$   & 20.0 & Si XIII(i or r) & 1854 or 1865\\ [1.0ex]
\hline
\enddata

\tablenotetext{a}{Measured Flux in units of $10^{-6}$ photons cm$^{\rm{-2}}$ s$^{\rm{-1}}$} 
\tablenotetext{b}{Line widths, assuming the 
widths of the forbidden and intercombination lines are set to be equal for any single ion.}
\tablenotetext{c}{Improvement in $C$ to fit after adding emission line component}
\tablenotetext{d}{Forbidden, intercombination and resonance lines denoted by 
f, i and r. Expected Lab--frame line energies in parenthesis in eV.}
\tablenotetext{e}{Line detection significance at lower than 99\% confidence.}

\end{deluxetable}

\clearpage

\begin{deluxetable}{lccc}
\tabletypesize{\small}
\tablecaption{Summary of LETG Model parameters.}
\tablewidth{0pt}
\tablehead{
\colhead{Model Component} & \colhead{Fit Parameter} & \colhead{Value} 
& \colhead{$\Delta C$}}

\startdata

1. Power-law continuum & $\Gamma$ & $1.73^{+0.22}_{-0.19}$ \\
& normalization$^{a}$ & $3.0\pm0.8 \times 10^{-3}$ & \\ [0.5ex]

2. Scattered power-law & $\Gamma$ & 1.73 (tied) & 57.0 \\
& normalization$^{a}$ & $6.6\pm1.0 \times 10^{-5}$ \\ [0.5ex]

3. Galactic absorption & $N_{\mathrm{H}}^{Gal}$$^{b}$ & ($1.5^{+0.6}_{-0.5}) 
\times 10^{21}$  \\ [0.5ex]

4. Ionized reflection & $\log \xi_{\mathrm{refl}}$$^{c}$ & $<1.65$ 
& 96.5   \\
& normalization$^{a}$  & ($2.49^{+0.83}_{-0.62}) \times 10^{-5}$  \\ [0.5ex]

5. Photoionized emission & $N_{\rm H}$$^{b}$ & $10^{22}$$^{f}$ 
& 125.6  \\

& log $\xi$$^{c}$ & $1.82_{-0.33}^{+0.13}$ \\
& Mg abund  & $2.6^{+2.6}_{-1.5}$ \\
& Si abund & $6.5^{+9.95}_{-3.86}$  \\
& outflow velocity$^{d}$ & $+150^{+240}_{-210}$ \\
& normalization$^{e}$ ($k$) & ($2.4^{+0.7}_{-0.6}) \times 10^{-6}$  \\ [0.5ex]

\hspace{4mm} Second emission region  & log $\xi$$^{c}$  & $3.0\pm0.4$ & 10.0 \\
& normalization$^{e}$ ($k$) & $1.2^{+1.9}_{-0.9}\times 10^{-5}$ \\ [0.5ex]

6. Photoionized Absorber & $N_{\rm H}$$^{b}$ & ($1.85_{-0.11}^{+0.09}) \times 10^{23}$ & 383.5  \\
& log $\xi$$^{c}$ & $1.42_{-0.12}^{+0.20}$ \\ 
%& redshift (z)               & ($2.20^{+0.25}_{-0.15}) \times 10^{-2}$ \\
& blueshift & $-0.034\pm0.002c$ & 22.3 \\
& outflow velocity$^{d}$ & $-10200\pm600$ \\ [0.5ex]   

%\hspace{4mm}   Second absorption region & $N_{\mathrm{H}}^{2}$       & ($4.01^{+2.71}_{-1.76}) \times 10^{23}$                    & 6.0    \\
%                                        & log $\xi^{2}$              & $3.08^{+0.19}_{-0.16}$ ergs cm s$^{-1}$                    &        \\ [0.5ex]

7. Fit statistic & C & 498.4/450 d.o.f \\ [1ex]
\enddata

\tablenotetext{a}{Normalisation corresponds to flux measured at 1\,keV, 
in units of photons\,cm$^{-2}$\,s$^{-1}$\,keV$^{-1}$} 
\tablenotetext{b}{Column density in units of cm$^{-2}$.}
\tablenotetext{c}{Units of ionization parameter ($\xi$) in erg\,cm\,s$^{-1}$.}
\tablenotetext{d}{Units km\,s$^{-1}$.}
\tablenotetext{e}{Units of Xstar emission normalization factor ($k$) in units of
$10^{38}$\,erg\,s$^{-1}$\,kpc$^{-2}$.}
\tablenotetext{f}{Model parameter fixed in fit.}

%\end{tabular}
\end{deluxetable}

\end{document}